\documentclass[a4paper,11pt]{article}
\pdfoutput=1 

\usepackage[T1]{fontenc} 
\usepackage{amsmath,amssymb,mathtools}
\usepackage{color}
\usepackage{cite}
\usepackage[colorlinks=true,linkcolor=blue, citecolor=red, urlcolor=blue, bookmarks]{hyperref}
\usepackage{graphics}

\usepackage[text={17.1cm,24.6cm},centering]{geometry} 

\usepackage{multirow,makecell}
\usepackage{textcomp}
\usepackage{wasysym}
\numberwithin{equation}{section}

\newcommand{\beq}{\begin{equation}}
\newcommand{\eeq}{\end{equation}}
\newcommand{\bea}{\begin{eqnarray}}
\newcommand{\eea}{\end{eqnarray}}
\def\l{\lambda}

\begin{document} 

\title{\bf Boundary effects on symmetry resolved entanglement}

\author{Riccarda Bonsignori$^1$ and Pasquale Calabrese$^{1,2}$}

\maketitle
 \vspace{-10mm}
\begin{center}
{\it
$^{1}$ International School for Advanced Studies (SISSA) and INFN, Via Bonomea 265, 34136 Trieste, Italy\\\vspace{1mm}
$^{2}$International Centre for Theoretical Physics (ICTP), Strada Costiera 11, 34151 Trieste, Italy
}
\vspace{10mm}
\end{center}

\begin{abstract}
We study the symmetry resolved entanglement entropies in one-dimensional systems with boundaries. 
We provide some general results for conformal invariant theories and then move to a semi-infinite chain of free fermions. 
We consider both an interval starting from the boundary and away from it. 
We derive exact formulas for the charged and symmetry resolved entropies based on theorems and conjectures about 
the spectra of Toeplitz+Hankel matrices. 
En route to characterise the interval away from the boundary, we prove a general relation between the eigenvalues of Toeplitz+Hankel matrices  
and block Toeplitz ones. 
An important aspect is that the saddle-point approximation from charged to symmetry resolved entropies introduces algebraic
corrections to the scaling that are much more severe than in systems without boundaries. 

\end{abstract}

\baselineskip 18pt
\thispagestyle{empty}
\newpage

\tableofcontents

\section{Introduction}

The characterisation of the interplay between entanglement and internal symmetries has recently become the focus of an intense research 
activity \cite{GS,equi-sierra,bons,Luca,lr-14,cgs-18,fg-19,mdc-20b,fg-20,mdc-20,ccdm-20,tr-20,mrc-20,trac-20,Topology,Anyons,hc-20,as-20} 
aimed to have a deeper resolution of the structure of the reduced density matrix of many-body systems and quantum field theories. 
The theoretical work in this area has almost entirely focussed on systems with periodic boundary conditions (PBC). 
However, there are many fundamental reasons to investigate systems with boundaries, in particular open boundary conditions (OBC). 
Just to quote a few: experimental solid-state systems typically have OBC; in trapped cold atoms, the vanishing of the density beyond a
trapping length induces OBC in the inhomogeneous gas that can be treated with the methods of field theories in curved space \cite{dsvc-17};
in some non equilibrium situations such as a quantum quench, the initial state can be treated as a boundary state in imaginary time formalism \cite{cc-05}.  
The goal of this paper is to study the effects of boundary conditions, in particular open, on symmetry resolved entanglement in conformal field theories and in free fermions chains.
The latter can be treated with simple exact methods.

The focus of our work is the \textit{entanglement entropy}.
Given a quantum system described by a pure state density matrix $\rho=|\psi \rangle  \langle \psi |$ and a bipartition of the Hilbert space 
$\mathcal{H}= \mathcal{H}_A \otimes \mathcal{H}_B$, the entanglement entropy is the Von Neumann entropy of the reduced density matrix $\rho_A=\mbox{Tr}_B\rho$, i.e.
\begin{equation}
S_{\rm vN}=- \mbox{Tr} \rho_A \log \rho_A,
\end{equation}
which is the limit for $n\rightarrow 1$ of a larger family of entropies, known as \textit{Renyi entropies}
\begin{equation}
S_n=\frac{1}{1-n} \log \mbox{Tr}(\rho_A^n).
\end{equation}
The latter give more information than the entanglement entropy, since their knowledge for different $n$ provides the full spectrum of the reduced 
density matrix $\rho_A$ \cite{cl-08}.

 \begin{figure}[t]
\centering
  \includegraphics[width=0.9\linewidth]{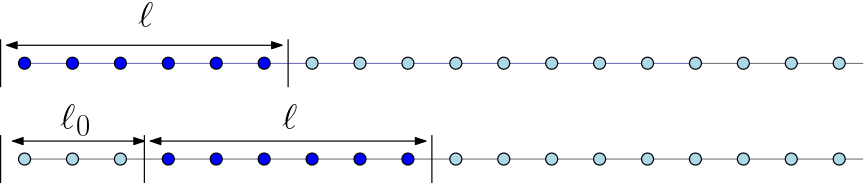}
  \caption{The two geometries we consider in this manuscript for a semi-infinite system. 
  The subsystem $A$ is always an interval of length $\ell$ that starts either from the boundary (top) or at distance $\ell_0$ from it (bottom with $\ell_0=3$).}
 \label{fig1}
\end{figure}

One of the remarkable results about the (R\'enyi) entanglement entropy in extended quantum systems is its universal scaling behaviour in conformal field theory (CFT)
when the subsystem A is an interval of length  $\ell$ embedded in the infinite line \cite{cc-04,cc-09,hlw-94,vlrk-03}
\begin{equation}
 S_n=\frac{c}{6}\left(1+\frac{1}{n}\right)\log {\ell}+c'_n,
 \label{Spbc}
 \end{equation}
where $c$ is the central charge and $c'_n$ is a non-universal additive constant. 
The presence of a boundary strongly affects the above scaling behaviour. 
For example, for a semi-infinite boundary CFT starting (say) at $x=0$, the scaling of the Renyi entropies for the interval $A=[0,\ell]$ (see Fig. \ref{fig1}) is \cite{cc-04,cc-09}:
\begin{equation}
S_n=\frac{c}{12}\left( 1+\frac{1}{n}  \right)\log \ell+\tilde{c}_n',
\label{Sobc}
\end{equation}
where the non universal constants $\tilde{c}_n'$ are related to the $c_n'$ in Eq. \eqref{Spbc} by the universal relation \cite{cc-04, zbca-06}
\begin{equation}
\tilde{c}_n'-\frac{c_n'}{2}= \log g,
\label{cvsc}
\end{equation}
and $\log g$ is the boundary entropy \cite{al-91,cs-17}. For the tight-binding chain with OBC that we are going to consider, we have  $g=1$\cite{al-91}. 
In a microscopic gapless model (e.g., a spin chain) the leading corrections to the asymptotic conformal behaviour given by Eqs. \eqref{Spbc} and \eqref{Sobc}
decay as $\ell^{-2K/n}$ \cite{ce-10,ccen-10,cc-10,ot-15,ccp-10} and $\ell^{-K/n}$ \cite{lsca-06,fc-11,cc-10} for PBC and OBC, respectively
($K$ is the scaling dimension of a relevant operator as we will see later on). These corrections oscillate with $\ell$, except for PBC at $n=1$.

When the subsystem $A$ is placed at distance $\ell_0$ from the boundary (i.e. $A=[\ell_0,\ell_0+\ell]$, see Fig. \ref{fig1}), CFT predicts the general scaling form 
for the entanglement entropies
\begin{equation}
S_n= \frac{c}{12}\left( 1+\frac{1}{n}  \right)\log \frac{4 \ell^2 \ell_0 (\ell+\ell_0)}{(2 \ell_0+\ell)^2}+{c}'_n+ \frac{\log \tilde F_n(x)}{1-n},
\end{equation}
where $x\in [0,1]$ is the anharmonic ratio
\begin{equation}
x=\frac{\ell^2}{(2 \ell_0+\ell)^2},
\label{anhr}
\end{equation}
and $\tilde F_n(x)$ is a function depending on the full operator content of the considered boundary CFT and normalised as $\tilde F_n(0)=1$
(to recover the bulk result \eqref{Spbc} for $\ell_0\gg \ell $, i.e. $x\to 0$). 
Such a universal function should be calculated on a case by case basis, analogously to the one for two disjoint intervals in periodic systems \cite{fps-09,cct-09,atc-11}. 
Some general results for $\tilde F_n(x)$ in the boundary compact boson CFT (Luttinger liquid) appeared only very recently \cite{b-20,bds-20}.

In this work, we consider the symmetry resolved entanglement of CFT and free fermionic systems with boundaries. 
In Sec. \ref{3} we provide all definitions about symmetry resolved entanglement and review the necessary results from the literature. 
In Sec. \ref{Sec:cft} we present our results within boundary CFT. 
In Sec. \ref{4} we move to free fermions on the semi-infinite line and consider a block starting from the boundary which can be 
analysed with a generalisation of the Fisher-Hartwig formula. 
In order to study the case of a block away from the boundary, we first prove in Sec. \ref{sec:proof} a general relation 
between the spectra of certain Toeplitz+Hankel matrices (related to the case of interest) and block Toeplitz ones (related to two disjoint 
blocks in an infinite chain). 
Hence in Sec. \ref{Sec5} we calculate the charged entropy for two blocks in an infinite chain, which are then used in 
Sec. \ref{6} to infer the results for the interval away from the boundary. 
We conclude in Sec. \ref{7} with a summary of the results and further discussions.

\section{Symmetry resolved entanglement: CFT and free fermions}
\label{3}

We consider an extended quantum system possessing an internal $U(1)$ symmetry, generated by a local operator $Q$. 
The system is taken in a pure state described by a density matrix $\rho$ with a definite value of this conserved charge and hence $[\rho,Q]=0$.
We consider a bipartition in $A$ and $B$ and the charge operator $Q$ itself splits in the sum  $Q=Q_A\otimes {\mathbb I}_B+ {\mathbb I}_A\otimes Q_B$ 
of the charge operators $Q_A,Q_B$ associated to each subsystem. Consequently,  $[\rho_A,Q_A]=0$, implying that the reduced density matrix $\rho_A$ has a block diagonal form, 
in which each block corresponds to an eigenvalue $q$ of $Q_A$, i.e.
\begin{equation}
\label{rhoAblock}
\rho_A=\oplus_q \Pi_q \rho_A\Pi_q= \oplus_q [p(q)\rho_A(q)],
\end{equation}
where $\Pi_q$ is the projector on the eigenspace of the eigenvalue $q$ and $p(q)= \mbox{Tr}(\Pi_q \rho_A)$ is the probability that a measurement of $Q_A$ gives the eigenvalue $q$ as outcome. Each block $\rho_A(q)$ of the reduced density matrix is normalised so that $\mbox{Tr}\rho_A(q)=1$. 
The block decomposition of the reduced density matrix can be exploited to quantify the contributions of the different charge sectors to the total entanglement entropy. 
In fact, Eq.  \eqref{rhoAblock} allows us to rewrite the total entanglement entropy as \cite{nc-10,exp-lukin}
\begin{equation}
\label{decompositionSvN}
S_{\rm vN}=\sum_q p(q)S_{\rm vN}(q)-\sum_q p(q)\log(q)\equiv S^c+S^f,
\end{equation}
where we have introduced the \textit{symmetry resolved entanglement entropy} as the entanglement entropy associated to the block $\rho_A(q)$
\begin{equation}
\label{Svnqdef}
S_{\rm vN}(q)=-\mbox{Tr}[\rho_A(q)\log \rho_A(q)].
\end{equation}
The two terms in Eq. \eqref{decompositionSvN} are called  configurational entanglement entropy ($S^c$) \cite{exp-lukin,wv-03,bhd-18} and fluctuation (or number) entanglement entropy ($S^f$)\cite{exp-lukin,kusf}  respectively.  
The former measuring the total entropy due to each charge sector weighted with their probability and the latter the one due to the fluctuations of the charge within the 
subsystem $A$.
Similarly, we also define the \textit{symmetry resolved R\'enyi entropies} as
\begin{equation}
\label{Snqdef}
S_n(q)=\frac{1}{1-n}\log \mbox{Tr}[\rho_A(q)^n].
\end{equation}
The evaluation of the symmetry resolved R\'enyi and entanglement entropies from the previous definitions would require the knowledge of the resolution of the spectrum of $\rho_A$ in $Q_A$, which is not straightforward because of the nonlocal nature of the projector $\Pi_q$. 
An alternative route \cite{GS,equi-sierra} is based on the computation of the the  \textit{charged moments}
\begin{equation}
Z_n({\alpha})\equiv\mbox{Tr}[\rho_A^ne^{i\alpha Q_A}],
\end{equation}
whose Fourier transform
\begin{equation}
\mathcal{Z}_n(q)= \int_{-\pi}^{\pi}\frac{d \alpha}{2 \pi} e^{-i q \alpha} {Z}_n(\alpha)\equiv \mbox{Tr}[\Pi_q\rho_A^n],
\end{equation}
readily provides the symmetry resolved quantities in Eqs. \eqref{Svnqdef} and \eqref{Snqdef} as
\begin{equation}
S_n(q)=\frac{1}{1-n}\log \left[\frac{\mathcal{Z}_n(q)}{\mathcal{Z}_1(q)^n} \right], \quad \quad S_{\rm vN}(q)=-\partial_n \left[  \frac{\mathcal{Z}_n(q)}{\mathcal{Z}_1(q)^n} \right]_{n=1}.
\label{SvsZ}
\end{equation}
Notice that the probability $p(q)$ is nothing but $p(q)=\mathcal{Z}_1(q)$.
In the next subsection we show how  to evaluate the charge moments using the replica trick.

\subsection{Replicas and twist fields in CFT}

In the replica approach, the moments $ \mbox{Tr}\rho_A^n$ can be evaluated for any $(1+1)$-dimensional quantum field theory (QFT) as partition functions over  a suitable 
$n$-sheeted Riemann surface $\mathcal{R}_n$ in which the $n$ sheets (replicas) are cyclically joined along the subsystem $A$ \cite{cc-04,cc-09}. 
Similarly \cite{GS}, the charged moments find a geometrical interpretation by inserting an Aharonov-Bohm flux through such surface,  so that the total phase accumulated by 
the field upon going through the entire surface is $\alpha$. Then the partition function on such modified surface is nothing but the charged moments ${Z}_n(\alpha)$.

This partition function can be rewritten in terms of correlator of properly defined twist fields implementing twisted boundary conditions. 
Assuming, without loss of generality, that the Aharonov-Bohm flux is inserted between the  $j$-{th} and $(j+1)$-{th} replicas, we can write the action of the twist fields as \cite{GS}
\begin{equation}
\mathcal{T}_{n,\alpha}(x,\tau) \phi_i(x',\tau)= \begin{cases}\phi_{i+1}(x',\tau)e^{i \alpha \delta_{ij}}\mathcal{T}_{n,\alpha}(x,\tau), & \mbox{if } x<x', \\ 
\phi_i(x',\tau)\mathcal{T}_{n,\alpha}(x,\tau), & \mbox{otherwise.}  \end{cases}
\end{equation}
In terms of these composite twist fields, the charged moments for a single interval $A=[0,\ell]$ embedded in the infinite line are
\begin{equation}
Z_n(\alpha)=\langle  \mathcal{T}_{n,\alpha}(\ell, 0) \tilde{\mathcal{T}}_{n,\alpha}(0,0) \rangle,
\end{equation}
where $\tilde{\mathcal{T}}$ is the anti-twist field. 
In particular, in the case of a $(1+1)$-dimensional CFT, it has been shown that  these fields behave as primary operators with conformal dimension
\begin{equation}
\Delta_{n,\alpha}= \Delta_n+\frac{\Delta_{\alpha}}{n}, \quad \quad \quad 
\Delta_n=\frac{c}{12}\left( n-\frac{1}{n} \right),
\label{dimna}
\end{equation}
so that they can be written as $\mathcal{T}_{n,\alpha}=\mathcal{T}_n\mathcal{V}_{\alpha}$, where $\mathcal{T}_n$ are the standard twist fields 
with conformal dimension $\Delta_n$ and $\mathcal{V}_{\alpha}$, with conformal dimension $\Delta_{\alpha}$, is a field implementing the insertion of the Aharonov-Bohm flux.
It follows that ${Z}_n({\alpha})$ scales as
\begin{equation}
Z_n(\alpha)=c_{n,\alpha} \ell^{-\frac{c}{6}\left( n-\frac{1}{n} \right)-\frac{2\Delta_{\alpha}}{n}},
\label{Zna1}
\end{equation} 
where $c_{n,\alpha}$ is the normalisation constant of the composite twist field (with $c_{n,0}=c_n$). 
The previous arguments apply to a generic CFT. In the case of Luttinger liquid conformal field theories \cite{GS} (that are $c=1$ free compact scalar bosonic massless theories describing the universality class of many 1D critical systems of interest, such as free and interacting spin chains), the operator $\mathcal{V}_{\alpha}$ can be identified with the vertex operator $e^{i\alpha\varphi(z)}$, so that its conformal dimension is 
\begin{equation}
\Delta_{\alpha}=\left(  \frac{\alpha}{2 \pi}\right)^2K,
\end{equation}
where $K$ is the  Luttinger parameter, related to the compactification radius of the bosonic theory.

From the charged moments \eqref{Zna1}, we get the symmetry resolved moments $\mathcal{Z}_n(q)$ via Fourier transform, that in the limit of large $\ell$ is \cite{GS}
\begin{equation}
\label{SymmResMoments}
\mathcal{Z}_n(q)\simeq \ell^{-\frac{c}{6}\left(n-\frac{1}{n}  \right)}\sqrt{\frac{n \pi}{2 K\log \ell}}e^{\frac{n \pi^2(q-\bar q)^2}{2 K\log \ell}},
\end{equation}
where $\bar q\equiv \langle Q_A \rangle$ represents the average number of particles. The latter, being a non-universal quantity of the system, cannot be determined within CFT.

From Eq. \eqref{SymmResMoments} we straightforwardly read the leading order of the symmetry resolved (R\'enyi) entropy \eqref{SvsZ} as 
\begin{equation}
\label{Equipartition}
S_n(q)=S_n-\frac{1}{2}\log \left( \frac{2K}{\pi}\log \ell \right)+O(\ell^0), 
\end{equation}
where $S_n$ is the total entropy \eqref{Spbc}. 
We observe that at leading orders the symmetry resolved entanglement does not depend on $q$, that is, it has the same value in all the different charge sectors corresponding to different eigenvalues of the charge operator. This result is known as \textit{equipartition of entanglement} \cite{equi-sierra}. 

\subsection{Free fermions techniques}
\label{fftec}

The simplest lattice model described by a Luttinger liquid CFT (with $K=1$) is represented by free spinless fermions hopping on a 1D lattice, with Hamiltonian
\begin{equation}
\label{HamiltonianFermions}
H=-\sum_{l}c_l^{\dagger}c_{l+1}+c_{l+1}^{\dagger}c_l+2 h \left(c_l^{\dagger}c_l-\frac{1}{2}   \right),
\end{equation}
where the fermionic ladder operators $c_l$ obey canonical anti-commutation relations $\{c_l,c_m^{\dagger}\}=\delta_{l,m}$ and $h$ is the chemical potential. 
The Hamiltonian is straightforwardly diagonalised in the momentum space and its ground state is a Fermi sea with momentum $k_F=\arccos |h|$. 
The conserved $U(1)$ local charge of the model is given by $Q=\sum_i c_i^{\dagger}c_i$ and it can be split in the sum $Q=Q_A+Q_B$ for any spatial bipartition. 
The Jordan-Wigner transformation
\begin{equation}
c_l=\left( \prod_{m<l}\sigma_m^z \right)\frac{\sigma_l^x+i\sigma_l^y}{2},
\end{equation}
maps the Hamiltonian \eqref{HamiltonianFermions} to the XX spin chain
\begin{equation}
H=-\sum_{l}\frac{1}{2}[\sigma_l^x\sigma_{l+1}^x+\sigma_l^y\sigma_{l+1}^y]-h\sigma_l^z,
\end{equation}
where $\sigma_l^{x,y,z}$ are the Pauli matrices at site $l$. 
Depending on the boundary conditions, the sum over $l$ in the Hamiltonians can run over a finite, semi-infinite, or infinite number of sites.

The reduced density matrix for an arbitrary spatial subsystem can be obtained using Wick's theorem, and has the following form \cite{peschel2001,peschel2003,pe-09}
\begin{equation}
\rho_A=\det C_A \exp\left(  \sum_{j,l\in A}[\log(C_A^{-1}-1)]_{jl}c_j^{\dagger}c_l\right),
\label{rhoA}
\end{equation}
where $C_A$ is the correlation matrix, i.e. the matrix formed by the correlations  $\langle c_i^{\dagger}c_j\rangle$ with $i,j\in A$.
The fermionic reduced density matrix \eqref{rhoA} is equal to the spin reduced density matrix for the same subsystem $A$ only when $A$ is one interval (starting from the boundary if PBC are not imposed), 
because the Jordan-Wigner transformation is local within a compact subset. Conversely, for a non-compact bipartition $A\cup\bar A$ (such as for disjoint 
intervals with PBC or one interval away from the boundary in an open chain) the non-local nature of the Jordan-Wigner string makes the spin and fermion reduced density matrices different \cite{ip-09,atc-09,fc-10}.

Denoting with $|A|$ the total number of sites within $A$ (i.e. the length $\ell$ for a single interval), $C_A$ is a $|A|\times |A|$ real and symmetric matrix. 
This matrix can be diagonalised by an orthogonal transformation.  
We write the eigenvalues of $C_A$ as $(1+\nu_k)/2$, $k=1,\dots, |A|$. 
Exploiting the quadratic form of $\rho_A$ in Eq. \eqref{rhoA}, the charged moments are \cite{GS}
\begin{equation}
\label{LnChargedMoments}
Z_n(\alpha)=\prod_{i=1}^{|A|}\left[ \left(\frac{1+\nu_i}{2}   \right)^ne^{i\alpha}+ \left(\frac{1-\nu_i}{2}   \right)^n \right]= 
\exp \left({\sum_{i=1}^{|A|}f_n(\nu_i,\alpha)}\right),
\end{equation}
with 
\beq
 f_n(x,\alpha)=\log\left[\left(\frac{1+\nu_i}{2}   \right)^ne^{i\alpha}+ \left(\frac{1-\nu_i}{2}   \right)^n \right].
 \label{fnxa}
\eeq
This formula can be straightforwardly used to evaluate the charged moments numerically in any Gaussian state and for any bipartition. 
For $\alpha=0$, we get back the standard formula for the (neutral) total moments. 

Eq. \eqref{LnChargedMoments} also is the starting point for the analytic computation of $Z_n(\alpha)$ using the Fisher-Hartwig formula, 
as done in Ref. \cite{bons}.
To this aim, one first introduces the determinant \cite{JK}
\begin{equation}
\label{detG}
D_{A}(\lambda)=\det[(\lambda+1)I-2C_A]=\prod_{j=1}^{|A|}(\lambda-\nu_j)\equiv \det(G),
\end{equation}
which is a polynomial of degree $\ell$ in $\lambda$ whose zeroes are the eigenvalues $\{\nu_j, j=1,\cdots,\ell\}$.
Then we rewrite Eq. \eqref{LnChargedMoments} in integral form
\begin{equation}
\label{integralLnZ}
\log Z_n(\alpha)=\frac{1}{2\pi i}\oint d \lambda f_n(\lambda,\alpha)\frac{d \log D_A(\lambda)}{d \lambda},
\end{equation}
where the contour integral encircles the segment $[-1,1]$ which is the support of the $\nu_j$.

So far, everything was completely general and applies to an arbitrary Gaussian state for an arbitrary spatial subsystem $A$. 
There is even no reference to the boundary conditions. 
In the following we review some exact results valid for the ground state of an infinite chain, focusing on those aspects we will need 
for the generalisation to open chains. 

\subsubsection{Exact results for the infinite chain}

We now specialise to the ground state of the infinite free-fermion chain and for the case of $A$ being an interval of length $\ell$.
The reduced correlation matrix is 
\begin{equation}
(C_A)_{ij}=\frac{\sin\left(k_F(i-j)\right)}{\pi(i-j)},
\label{CApbc}
\end{equation}
with $i,j=1\dots \ell$. The matrix $G$ in Eq. \eqref{detG} has a Toeplitz form, i.e. its elements depend only on the difference between row and column indices $G_{jk}=g_{j-k}$.
For this matrix, the integral \eqref{integralLnZ} can be evaluated analytically in the asymptotic limit $\ell\to \infty$ using the Fisher-Hartwig formula \cite{basor},
a technique that has been used extensively to evaluate entanglement in free lattice models \cite{JK,ce-10,km-05,afc-09,fik-07,fc-11,aef-14,aef-14b,aef-15}. 
We briefly recap this derivation in the following in order to illustrate the procedure and set the notation that will be used for open systems.

The Fisher-Hartwig formula is written in terms of the \textit{symbol} of the Toeplitz matrix, defined as the Fourier transform of $g_l$
\begin{equation}
\label{Toeplitzelement}
g_l=\int_{-\pi}^{\pi}\frac{d \theta}{2 \pi} e^{i l\theta}g(\theta),
\end{equation}
that in the case considered here is given by
\begin{equation}
\label{symbol}
g(\theta)=\left\{
                \begin{array}{ll}
                  \lambda+1,\quad \theta \in [-\pi , -k_F] \cup [k_F,\pi],\\
                 \lambda-1,\quad \theta \in [-k_F,k_F].
                \end{array}
              \right.
\end{equation}
In the integration domain \eqref{Toeplitzelement} the symbol has two discontinuities, located at $\theta_1=-k_F$ and $\theta_2=k_F$. 
The Fisher-Hartwig formula relies on the possibility to express the symbol in the following form
\begin{equation}
\label{SymbolFH}
g(\theta)=f(\theta)\prod_{r=1}^{R}e^{ib_r[\theta-\theta_r-\pi\, {\rm sgn}(\theta-\theta_r)]}(2-2\cos(\theta-\theta_r))^{a_r},
\end{equation}
where $R$ is an integer, $a_r,b_r,\theta_r$ are constants and $f(\theta)$ is a smooth function with winding number zero. 
For the symbol $g(\theta)$ in Eq. \eqref{symbol}, there are two discontinuities so that $R=2$, and the constants assume the values 
$a_{1,2}=0$, $b_2=-b_1=\beta_{\lambda}+m$ and $f(\theta)=f_0=(\lambda+1)e^{-2ik_Fm}e^{-2ik_F\beta_{\lambda}}$, where
\begin{equation}
\label{Beta}
\beta_{\lambda} = \frac{1}{2 \pi i} \log \left[ \frac{\lambda+1}{\lambda-1} \right],\qquad {\rm with}\quad
\frac{d \beta_\lambda}{d \lambda}=\frac{1}{\pi i}\frac1{1-\lambda^2},
\end{equation}
and the integer $m\in {\mathbb Z}$ labels the different inequivalent representations of the symbol.
Usually one refers simply to the Fisher-Hartwig formula when there is a single representation of the symbol and to the generalised one when 
there are multiple representations, as it is the case for us. 
For a Toeplitz matrix $T$ with a symbol of the form \eqref{symbol} without inequivalent representation, the Fisher-Hartwig formula provides the large $\ell$ behaviour
\begin{equation}
\det T \simeq F[f(\theta)]^{\ell}\left( \prod_{j=1}^R\ell^{a_j^2-b_j^2} \right), \qquad
{\rm where} \;\;  F[f(\theta)]=\exp\left( \frac{1}{2\pi} \int_0^{2\pi}d \theta \log f(\theta)\right).
\eeq 
When the symbol has several inequivalent representations, as for our case, one has to sum over all of them \cite{basor,ce-10}, obtaining, in our specific case,
the asymptotic for large $\ell$ 
\begin{equation}
\label{GeneralizedFH}
D_A(\lambda)\simeq (\lambda+1)^{\ell}\left(\frac{\lambda+1}{\lambda-1}\right)^{-\frac{k_F \ell}{\pi}}\sum_{m \in \mathbb{Z}}(2 \ell |\sin k_F|)^{-2(m+\beta_{\lambda})^2}
e^{-2ik_F m \ell}\left[ G(m+1+\beta_{\lambda})G(1-m-\beta_{\lambda}) \right]^2,
\end{equation}
where $G (z)$ is the Barnes $G$-function.

The charged moments $Z_n(\alpha)$ are evaluated by inserting the result for $D_A(\lambda)$ \eqref{GeneralizedFH} into the integral \eqref{integralLnZ}.
It is easy to see that, for $\alpha\in [-\pi,\pi]$, the leading behaviour for large $\ell$ of such integral is given by the term with $m=0$
\begin{equation}
D_A^{(0)}(\lambda)\equiv (\lambda+1)^{\ell}\left(\frac{\lambda+1}{\lambda-1}\right)^{-\frac{k_F \ell}{\pi}}
(2 \ell |\sin k_F|)^{-2\beta_{\lambda}^2}\left[ G(1+\beta_{\lambda})G(1-\beta_{\lambda})\right]^2,
\end{equation}
and so the contour integral \eqref{integralLnZ} gives \cite{bons}
\begin{equation}
\label{LnZPBC}
\log Z_n^{(0)}(\alpha)=  i \alpha \frac{k_F \ell}{\pi}-
\left[ \frac{1}{6}\left( n-\frac{1}{n} \right) + \frac{2}{n}\left( \frac{\alpha}{2 \pi} \right)^2\right] \log(2 \ell |\sin k_F|) +   \Upsilon{(n,\alpha)} .
\end{equation}
In this expression for the charged moment, we recognise immediately the average number of particles (the linear term in $\alpha$) given by 
$\bar q\equiv \langle Q_A \rangle=k_F\ell/\pi$, and the dimension of the modified twist field from the 
term proportional to $\log(2 \ell |\sin k_F|)$. The non-universal constant $\Upsilon(n,\alpha)$ is given by the integral
\begin{equation}
\label{Upsilon}
\Upsilon{(n,\alpha)}= {n i}\int_{-\infty}^\infty  dw [\tanh(\pi w)-\tanh (\pi n w+i\alpha/2)]  \log \frac{\Gamma(\frac12 +iw)}{\Gamma(\frac12 -iw)} \,,
\end{equation}
that is real as long as $\alpha$ is real.
For later purposes, we rewrite it as
\begin{equation}
\label{UpsilonExpansion}
\Upsilon (n, \alpha) = \Upsilon(n) + \gamma_2 (n) \alpha^2 + \epsilon (n, \alpha), \qquad \epsilon (n, \alpha)= O (\alpha^4),
\end{equation}
where
\begin{equation}
\gamma_2{(n)}= \frac{n i}4\int_{-\infty}^\infty  dw [\tanh^3(\pi nw)-\tanh (\pi n w)]  \log \frac{\Gamma(\frac12 +iw)}{\Gamma(\frac12 -iw)} \,.
\label{gamma2}
\end{equation}
For $\alpha=0$, the above results reproduce the well known (total) R\'enyi entropies \cite{JK,ce-10}. 

The symmetry resolved moments are just the Fourier transform $\mathcal{Z}_n(q)$ of ${Z}_n(\alpha)$. 
In this Fourier transform we ultimately use a saddle-point approximation in which $Z_n(\alpha)$ is Gaussian and hence 
we truncate hereafter $Z^{(0)}_n(\alpha)$ in Eq. (\ref{LnZPBC}) at quadratic order in $\alpha$.
Consequently, the charged partition function can be well approximated as
\begin{equation}
Z_n(\alpha)= Z_n(0)  e^{i \alpha \bar q- b_n \alpha^2/2}, 
\label{ZG}
\end{equation}
where 
\begin{equation}
b_n= \frac{b}{\pi^2 n}\ln \ell  -h_n, 
\label{bn}
\end{equation}
with
\begin{equation}
b=1, \qquad h_n\equiv -\frac{1}{\pi^2 n}\ln (2 |\sin k_F|) +2 \gamma_2(n) . 
\label{bn2}
\end{equation}
(Here we slightly change our notations compared to Ref. \cite{bons} and follow more closely those in \cite{mdc-20b}.)
In this way, the Fourier transform is a simple Gaussian integral, with result 
\begin{equation}
\label{eq:SP-FTrans-step3}
\mathcal{Z}_n(q)=\frac{Z_n(0)}{\sqrt{2\pi b_n}}e^{-\frac{(q-\bar q)^2}{2 b_n}}.
\end{equation}
The symmetry resolved R\'enyi entanglement entropies are then obtained using Eq. \eqref{SvsZ} as
\begin{equation}
\label{eq:SP-SRRE}
S_n(q)=
S_n-\frac12\ln(2\pi)+\frac{1}{1-n}\ln\frac{b_1(\ell,t)^{n/2}}{b_n(\ell,t)^{1/2}}-\frac{q^2}{2(1-n)}\left( \frac{1}{b_n(\ell,t)}-\frac{n}{b_1(\ell,t)} \right), 
\end{equation}
where $S_n$ is the total entropy. 
Expanding for large $\ell$, we have
\begin{multline}
\label{eq:SP-SRRE-v2Order}
S_n(q)=
S_n-\frac{1}{2} \ln \Big(\frac{2b}{\pi} \ln \delta_n \ell \Big)+ \frac{\ln n}{2(1-n)}-\frac{\pi^4n(h_1-nh_n)^2}{4(1-n)^2(b\ln \ell)^2}+ 
\\ +
(q-\bar q)^2n \pi^4\frac{h_1-nh_n}{2(1-n)(b\ln \ell \kappa_n)^2} + o(\ln \ell^{-2}),
\end{multline}
where, following Ref. \cite{bons} we absorbed some subleading corrections in the amplitudes as
\begin{equation}
\label{eq:deltan}
\ln \delta_n=-\dfrac{\pi^2 n  (h_n-h_1)}{b(1-n)},
\qquad 
\ln \kappa_n=-\pi^2\frac{(h_1+n h_n)}{2b}.
\end{equation}
The above formula is valid also for the symmetry resolved Von Neumann entropy taking properly the limits of the various pieces as $n\to 1$. 

We wrote these formulas in a rather generic fashion as a function of $b_n$ because in the other cases studied here (and 
elsewhere \cite{mdc-20b,Luca,mrc-20}) only the specific form of $b_n$ (or $h_n$) matters and the final result is always given by 
Eq. \eqref{eq:SP-SRRE-v2Order} with the minor redefinition of the amplitudes.

\section{Entanglement entropy, CFT, and boundaries}
\label{Sec:cft}

In this section we present our boundary CFT results for the charged and symmetry resolved entropies. 
We start from  a 1D system in a semi-infinite line $[0;\infty)$ and a subsystem A consisting in a finite interval $[0;\ell)$ as in Fig. \ref{fig1}. 
From general CFT scaling, one expects the charged moments ${Z}_n(\alpha)$ to be 
\begin{equation}
{Z}_n(\alpha)\equiv {\rm Tr}\left[ \rho_A^n e^{i Q_A \alpha}  \right]= \langle {\cal T}_{n,\alpha} (\ell)\rangle_{\rm HP}=
 \tilde c_{n,\alpha}( 2\ell)^{-\frac{c}{12}(  n-\frac{1}{n} )-\frac{\Delta_{\alpha}}{n}},
 \label{Znabcft}
\end{equation}
where the subscript HP stands for the average over the (right) half plane $z=x+i \tau$ with $x\in {\mathbb R}^+$ and $\tau\in {\mathbb R}$. 
Anyhow, it is worth to obtain such (correct) prediction from first principles by merging the boundary CFT approach to the entanglement entropy \cite{cc-04,cc-09} 
with the insertion of the Aharonov-Bohm flux \cite{GS}. 

The $n$-sheeted Riemann surface then consists of $n$ copies of the half plane $x\geq 0$ sewn together along $0 \leq x\leq \ell, \tau=0$. 
Once again, the flux between the $j$-{th} and $(j+1)$-{th} replicas can be implemented by the definition of local composite twist fields 
$\mathcal{T}_{n,\alpha}=\mathcal{T}_n\mathcal{V}_{\alpha}$ 
at the end-point of the region $A$, where $\mathcal{V}_{\alpha}$ is responsible for the Aharonov-Bohm flux and $\mathcal{T}_n$ generates the Riemann geometry.
The scaling dimension of the composite twist field is obtained by evaluating the expectation value of the total stress-energy tensor $T(w)=\sum_{j=1}^nT_j (w)$ in 
the Riemann geometry $\mathcal{R}_n$ with inserted flux $\alpha$. First, the transformation 
\begin{equation}
z=\left(\frac{\ell-w}{\ell+w}\right)^{1/n} ,
\label{map}
\end{equation}
maps the whole Riemann surface into the unit disc $\mathcal{D}=\{ |z|<1\}$ with the flux $\alpha$.
Thus we have
\begin{equation}
\langle  T(w) \rangle_{\mathcal{R}_{n,\alpha}}=\sum_j\left( \frac{d z}{d w} \right)^2 \langle T_j(z) \rangle_{\mathcal{D},\alpha}+\frac{n c}{12}\{z,w\},
\label{TT}
\end{equation}
where the Schwartzian derivative is given in this case by
\begin{equation}
\frac{c}{12}\{z,w\}=\frac{c}{24}\left(1-\frac{1}{n^2} \right)\frac{(2 \ell)^2}{(w- \ell)^2(w+\ell)^2} .
\label{sch}
\end{equation}
In Eq. \eqref{TT} we added the subscript $\alpha$ to stress that the expectation values are taken in the presence of a flux.
Notice that $ \langle T_j(z) \rangle_{\mathcal{D},\alpha}$ is non-zero for $\alpha\neq 0$. 
To calculate it, let us start by noticing that the transformation \eqref{map} maps the subsystem $A$ (say on the first sheet) into $[0,1]$ and the branch point into the origin.
Hence, a closed path encircling the branch point $n$ times is mapped into a single-winding orbit around the origin so that 
\beq
 \langle T_j(z) \rangle_{\mathcal{D},\alpha}= \frac{ \langle T_j(z) \mathcal{V}_{\alpha}(0) \rangle_\mathcal{D}}{\langle \mathcal{V}_{\alpha}(0)\rangle_\mathcal{D}},
\label{p2p2}
\eeq
where the subscript ${\cal D}$ (without $\alpha$) refers to the expectation values on the unit disk in the absence of the flux. 
When ${\cal V}_\alpha$ is a primary operator, the right hand side of Eq. \eqref{p2p2} is ${h_\alpha}/{z^2}$ \cite{cardy-84} and hence
\beq
 \langle T_j(z) \rangle_{\mathcal{D},\alpha}=\frac{h_\alpha}{z^2}\,.
\eeq
Plugging the above expression and the Schwartzian derivative \eqref{sch} into Eq. \eqref{TT} ones get
\begin{equation}
\frac{\langle  T(w)\mathcal{T}_{n,\alpha}(\ell) \rangle_{\rm HP}}{\langle \mathcal{T}_{n,\alpha}( \ell) \rangle_{\rm HP}}=
\left[ \frac{c}{24}\left(  n-\frac{1}{n} \right)+\frac{h_{\alpha}}{n} \right]\frac{(2 \ell)^2}{(w-\ell)^2(w+ \ell)^2}.
\label{pp4}
\end{equation}
The comparison of Eq. \eqref{pp4} with the conformal Ward identity for boundary CFT \cite{cardy-84} confirms that also in the presence of boundaries, 
the scaling dimension of the composite twist fields is \eqref{dimna}, leading immediately to the expected result \eqref{Znabcft}. 
Taking the Fourier transform by saddle point approximation, one obtains the asymptotic symmetry resolved moments
\begin{equation}
\label{ZnqCFTboundary}
\mathcal{Z}_n(q)\simeq (2 \ell)^{-\frac{c}{12}\left(n-\frac{1}{n} \right)}\sqrt{\frac{n\pi}{K\log \ell}}e^{\frac{n\pi^2(q-\bar q)^2}{K\log\ell}}.
\end{equation}

The other situation of interest is that of an interval of length $\ell$ placed at distance $\ell_0$ from the boundary.
In this case, global conformal invariance fixes the overall scaling. 
Indeed, the charged moment is a two-point function $Z_n(\alpha)=\langle {\cal T}_{n,\alpha}(u_1)\tilde {\cal T}_{n,\alpha} (v_1)\rangle_{\rm HP}$ 
with $u_1=\ell_0$ and $v_1=\ell+\ell_0$. 
By image charges technique, this is related to a four-point function on the plane with images at $u_2=-v_1$ and $v_2=-u_1$.
The scaling of a general four four-point function of composite twist fields is  
\begin{equation}
\label{TwoInter}
\langle {\cal T}_{n,\alpha}(u_2) \tilde {\cal T}_{n,\alpha}(v_2) {\cal T}_{n,\alpha}(u_1) \tilde {\cal T}_{n,\alpha}(v_1) \rangle=
c_{n, \alpha}^2\left( \frac{|u_1-u_2||v_1-v_2|}{|u_1-v_1||u_2-v_2||u_1-v_2||u_2-v_1|} \right)^{2\Delta_{n,\alpha}}F_{n,\alpha}(x),
\end{equation}
where $F_{n,\alpha}(x)$ is a universal function that depends on the operator content of the CFT and  $x$ is the anharmonic ratio of the four points.
The desired boundary correlation is related to the square root of the bulk four-point function, and hence, 
specifying to the actual values of $u_i$ and $v_i$, we have 
\begin{equation}
{\rm Tr}\left[ \rho^n_A e^{i Q_A\alpha}  \right]= \langle {\cal T}_{n,\alpha}(u_1)\tilde {\cal T}_{n,\alpha} (v_1)\rangle_{\rm HP}= 
c_{n,\alpha}\left( \frac{(2 \ell_0+\ell)^2}{4\ell^2\ell_0(\ell+\ell_0)} \right)^{\Delta_{n,\alpha}}\tilde F_{n,\alpha}(x),
\label{Sadet}
\end{equation}
where $\tilde F_{n,\alpha}(x)$ is a universal function of the anharmonic ratio $x$ in \eqref{anhr} that depends on the operator content of the {\it boundary} CFT.
As $\ell_0\gg \ell$, using $\tilde F_{n,\alpha}(0)=1$, Eq. \eqref{Sadet} tends to the results for one interval in an infinite system \eqref{Zna1}, as it should. 
Conversely, for $\ell_0\ll \ell$, one recovers the result for the interval close to the boundary, cf. Eq. \eqref{Znabcft}, in which the relation between 
$c_{n,\alpha}$ and $\tilde c_{n,\alpha}$ (generalisation to $\alpha\neq 0$ of Eq. \eqref{cvsc}) 
\beq
\frac{\tilde c_{n,\alpha}}{c_{n,\alpha}^{1/2}}= g^{1-n},
\label{cvsca}
\eeq
is provided by the singular behaviour of $F_{n,\alpha}(x)$ close to $x=1$.

Finally, let us briefly mention what happens for a finite system of length $L$ with OBC on both sides (which is the most relevant situation for physical
applications, as e.g., those in Refs. \cite{dsvc-17,cc-05}). The worldsheet of each replica is an infinite (in the time direction) strip of length $L$. This can be mapped to the 
half plane by a conformal transformation (the logarithm). Hence the net effect of having a finite system is just to replace all separations in Eqs.
\eqref{Znabcft} and \eqref{Sadet} with the appropriate chord distance, e.g. in Eq. \eqref{Znabcft} we have the replacement
\begin{equation}
\ell\to \frac{L}\pi \sin\frac{\pi \ell}L,
\end{equation}
and similarly in Eq. \eqref{Sadet}.
In the same way, one can consider a semi-infinite system at finite temperature, with the worldsheet being a semi-infinite cylinder that can be mapped to the half plane
by a conformal transformation. Simple algebra leads to the replacement $2\ell\to \frac{\beta}\pi \sinh\frac{2\pi \ell}\beta$ \cite{cc-04,cc-09}, 
with $\beta$ being the inverse temperature.

\section{Semi-infinite chain of free fermions: the block $A=[1,\ell]$}
\label{4} 

The results in the previous section are valid for an arbitrary boundary CFT.
In this section, we specialise to a particular microscopic model whose low energy physics is captured by a CFT. 
We wish to calculate also the non-universal factors (not predicted by CFT) that enter in a parameter-free comparison with numerics. 
We focus on the tight-binding model given by the Hamiltonian \eqref{HamiltonianFermions}, which describes free fermions hopping on a 1D 
lattice. We work with a semi-infinite chain, i.e. $l\in {\mathbb N}$, and the first site on the left is free, i.e. we choose open boundary conditions. 

The fermion correlation function between two arbitrary sites $i,j$ \cite{peschel2001} is 
\begin{equation}
(C_A)_{ij}=\frac{\sin\left(k_F(i-j)\right)}{\pi(i-j)}-\frac{\sin\left(k_F(i+j) \right)}{\pi (i+j)}.
\label{Cij}
\end{equation}
For an  arbitrary spatial bipartition $A\cup\bar A$, the elements of the correlation matrix $C_A$ are just given by the above correlation with indices $i,j$ restricted in $A$. 
The first term in Eq. \eqref{Cij} is the same as in the infinite system (cf. Eq. \eqref{CApbc}) and it depends only on the spatial distance between 
the two points as a consequence of translational invariance. 
The second term breaks the latter symmetry and depends on the average distance from the boundary (and it vanishes for large $i,j$ at fixed separation $|i-j|$, when only 
the first term survives).
 
Having constructed the subsystem correlation matrix, the entanglement spectrum and hence the R\'enyi entropies (total and symmetry resolved) follow from the 
general results in Sec. \ref{fftec}. In particular, the charged moments are given by Eq. \eqref{LnChargedMoments} in terms of the eigenvalues $\nu_j$ of the matrix $C_A$. 
In the following we specialise to two bipartitions that can be handled analytically, which are the ones depicted in Fig. \ref{fig1} with one block of $\ell$ sites 
at distance $\ell_0$ from the boundary (that can be $\ell_0=0$, recovering an interval starting from the boundary).
The $\ell\times \ell$ correlation matrix may be written as 
\beq
(C_A)_{ij}=f_{i-j}- f_{i+j+2\ell_0}, \qquad i,j= 1, \dots, \ell,\qquad \ell_0=0,1, 2, \dots
\label{T+H}
\eeq
with $f_{k}$ the elements of the correlation matrix for the infinite system in Eq. \eqref{CApbc}. 
The first term in the rhs of Eq. \eqref{T+H} is a Toeplitz part, since it depends only on the difference of the indices, and is equal to the matrix for an infinite system; 
the second term is a Hankel matrix since it depends only on the {\it sum} of the indices and it comes from the presence of the boundary.
The correlation matrix $C_A$ in Eq. \eqref{T+H} is then equal to the sum of one Toeplitz matrix and a Hankel one.

\subsection{A generalised Fisher-Hartwig formula and charged entropies}

In the case of a block starting from the boundary, i.e. with $\ell_0=1$, the Toeplitz+Hankel matrix in Eq. \eqref{T+H} has been studied a lot in both the mathematical 
and physics literature \cite{dik-10,fc-11,be-17}. 
The final result of interest for our paper is the conjectured form for a generalised Fisher-Hartwig expansion \cite{fc-11}
\begin{multline}
D_A(\l)\simeq  (\l+1)^\ell\left(\frac{\l+1}{\l-1}\right)^{-\frac{k_F\ell}\pi}
\sum_{m\in\mathbb{Z}}e^{i\frac{\pi}{2}(\beta+m)}
\left[ 4\Bigl(\ell+\frac12\Bigr) |\sin k_F|\right]^{-(m+\beta_\l)^2}  \\
\times e^{-2ik_F(\beta+m)(\ell+{1}/{2})} G(m+1+\beta_\l)G(1-m-\beta_\l),
\label{Drep2o}
\end{multline}
where $\beta_{\lambda}$ is defined in Eq. \eqref{Beta} and $G(z)$ is the Barnes function. 
This formula was conjectured in Ref. \cite{fc-11} in such a way to reproduce the results for $m=0$ derived rigorously in Ref. \cite{dik-10} and 
generalising heuristically to the existence of inequivalent representations of the symbol. 
The term $1/2$ added to $\ell$ has been introduced to absorb some of the subleading corrections to better match numerical evaluation of the determinant and 
has not a rigorous basis (we will see in the following section another reason for its presence). 
This asymptotic expansion has been used to compute the moments of the reduced density matrix $\mbox{Tr}\rho_A^n$, for which 
the leading term corresponds to $m=0$, while the first corrections are given by $m=\pm 1$. 
For the charged moments $Z_n(\alpha)$, restricting by periodicity to $\alpha\in[-\pi,\pi]$, the leading term is still given by $m=0$, 
see also the discussion in Ref. \cite{fg-20}.

\subsubsection{The leading term}

The leading behaviour of the determinant $D_A(\lambda)$ is given by
\begin{equation}
D_A^{(0)}(\lambda) \sim e^{i \left( \frac{\pi}{2}-k_F \right)\beta_{\lambda}}  \left[  (\lambda+1)  \left( \frac{\lambda+1}{\lambda-1}  \right)^{-\frac{k_F}{\pi}} \right]^{\ell}
\left( 4 \ell |\sin k_F| \right)^{-\beta_{\lambda}^2}G(1-\beta_{\lambda})G(1+\beta_{\lambda}),
\label{DA0}
\end{equation}
(at the leading order we can drop the $1/2$ in the generalised form)
so that the leading contribution to the charged moments \eqref{integralLnZ} is
\begin{equation}
\label{lnZ0}
\log {Z}^{(0)}_n(\alpha)=\frac{1}{2 \pi i} \oint d \lambda f_{n,\alpha}(1,\lambda)\frac{d \log D^{(0)}_{\ell}(\lambda)}{d \lambda}=a_0+a_1 \ell+a_2 \log L_k+ a_3,
\end{equation}
where
\begin{equation}
\begin{split}
a_0&=\frac{1}{2\pi } \left( \frac{\pi}{2}-k_F \right)\oint d \lambda f_{n,\alpha}(1,x)\frac{d}{d \lambda}\beta_{\lambda},\\
a_1&= \frac{1}{2 \pi i} \oint d \lambda f_{n,\alpha}(1,\lambda)\left( \frac{1-k_F/\pi}{1+\lambda}- \frac{k_F/\pi}{1-\lambda} \right),\\
a_2&=\frac{1}{2 \pi i} \oint d \lambda f_{n,\alpha}(1,\lambda)\frac{d(-\beta_{\lambda}^2)}{d \lambda}= \frac{1}{\pi^2}
\oint d \lambda f_n(\lambda,\alpha) \frac{\beta_{\lambda}}{1-\lambda^2},\\
a_3&= \frac{1}{2 \pi i} \oint d \lambda f_{n,\alpha}(1,\lambda) \frac{d \log[G(1-\beta_{\lambda})G(1+\beta_{\lambda})]}{d \lambda}.
\end{split}
\end{equation}
Putting everything together, the final result is  
\begin{equation}
\label{lnZ0res}
\log {Z}_n^{(0)}(\alpha)=\frac{i \alpha k_F \ell}{\pi}+ \left[ \frac{i \alpha}{\pi}\left(\frac{k_F}{\pi} - \frac{1}{2}\right) \right]-\left[ \frac{1}{12} \left(n-\frac{1}{n}  \right)+\frac{1}{n}\left( \frac{\alpha}{2 \pi} \right)^2\right] \log\big[4 \ell  |\sin k_F|\big]+ \frac{\Upsilon(n,\alpha)}{2},
\end{equation}
where the real constant $\Upsilon(n,\alpha)$ is the same as the one in Eq. \eqref{Upsilon}.
Let us now discuss the various terms, emphasising the differences with the case of the infinite system in Eq. \eqref{LnZPBC}. 
The purely imaginary term gives the average number of particles in the interval $\ell$ which is 
\beq
\bar q\equiv\langle Q_A \rangle= \frac{k_F}{\pi} \ell+\frac1{\pi} \left(\frac{k_F}{\pi} - \frac{1}{2}\right) +o(1)\,.
\label{qq}
\eeq
The leading piece, proportional to $\ell$, is the average density also present in the infinite system. 
The $O(1)$ correction is the more interesting and represents the variation of the number due the inhomogeneity close to the boundary.
Indeed the density at the site $j$ is $n_j=k_F/\pi + \sin(2 k_F j)/(2 \pi j)$ (i.e. the correlation \eqref{Cij} at coinciding points).
Thus, the mean number of particles in $[1,\ell]$ is
\beq
\langle  Q_A \rangle -\frac{k_F}\pi \ell =\sum_{j=1}^\ell \frac{\sin(2 k_F j)}{2 \pi j}=\sum_{j=1}^\infty \frac{\sin(2 k_F j)}{2 \pi j}+O(\ell^{-1})= 
\frac1{\pi} \left(\frac{k_F}{\pi} - \frac{1}{2}\right)+o(1)\,.
\eeq
At half-filling this term is obviously zero since there is no inhomogeneity.  

Let us now move to the other pieces in Eq. \eqref{lnZ0res}.
The logarithmic term agrees with the boundary CFT prediction \eqref{Znabcft} for this specific model and confirms the dimension of the modified twist fields.
Interesting enough, also the non-universal constant is related to the one in the absence of boundaries and satisfies the universal relation \eqref{cvsca} with $g=1$, 
as well known.

\begin{figure}[t]
\includegraphics[width=.5\textwidth]{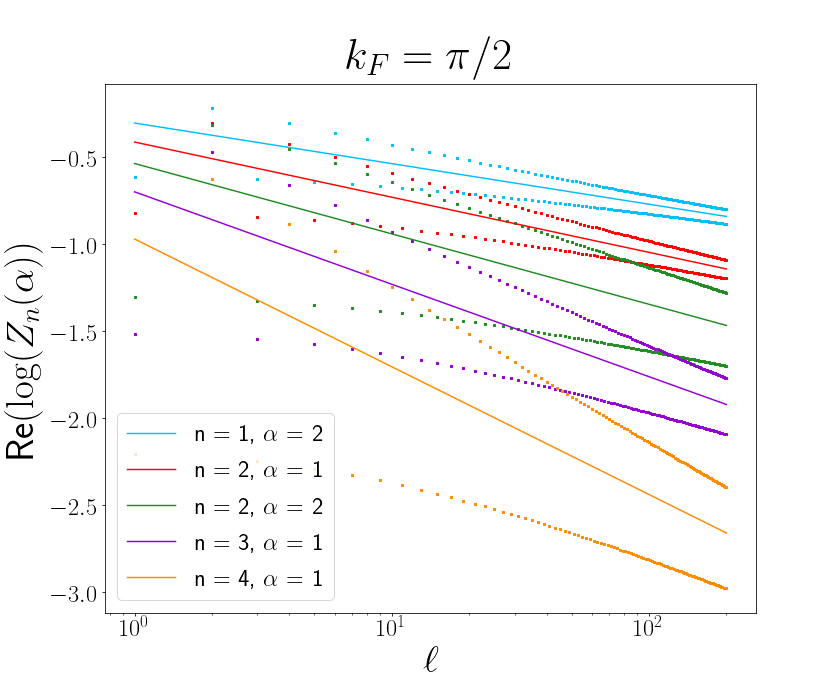}
\includegraphics[width=.5\textwidth]{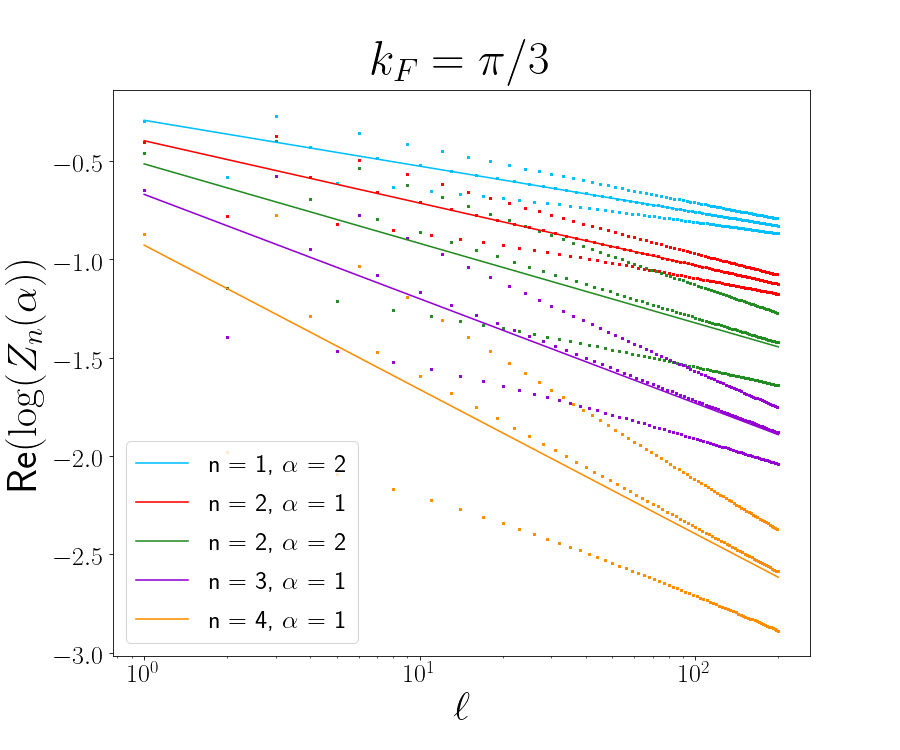} 
     \caption{Real part of the charged R\'enyi entropies $\log Z_n(\alpha)$ with the insertion of a flux $\alpha$ as functions of $\ell$.
     The exact numerical results (symbols) are shown for different values of $\alpha$ and $n$ at fillings $k_F=\pi/2$ (left) and $k_F=\pi/3$ (right). 
     The numerical data present a slow approach to the asymptotic behaviour \eqref{lnZ0res}, with large oscillating corrections to the scaling.
    }
\label{lnZFig}
\end{figure}

In Fig. \ref{lnZFig} we present the comparison between the analytical formula \eqref{lnZ0res} and the exact numerical data obtained for the charged R\'enyi entropies
for different values of $n$ and $\alpha$, at filling $k_F=\pi/2$ and $k_F=\pi/3$. 
It is evident that the data are slowly approaching the exact predictions, but there are large oscillating corrections to the scaling whose amplitude increases with $n$ and $\alpha$. 
These corrections do not come as a surprise: they have been intensively studied for the total entanglement entropy both for periodic \cite{ccen-10,ce-10,cc-10,ot-15,ccp-10} 
and open boundary conditions \cite{lsca-06,fc-11}. For  the charged moments with $\alpha\neq 0$ and PBC, they have been described and characterised in Refs. \cite{bons,fg-20}. 
In the following subsection we will derive and quantify them for OBC.

\subsubsection{The leading corrections to the scaling for the charged entropies}
The corrections to the scaling are encoded in the generalised Fisher-Hartwig formula \eqref{Drep2o}. 
The precise order in which they appear depends on the quantity to be calculated. 
It is straightforward to convince ourselves that for $0<\alpha<\pi$ ($-\pi<\alpha<0$) the leading correction is the one with $m=1$ ($m=-1$). 
For $\alpha=0$ the terms $m=\pm1$ are of the same order. If not yet clear, all this will be evident and self-consistent at the end of the calculation in this subsection.

We perform the calculation keeping only the two terms with $m=\pm1$ and approximate the characteristic polynomial $D_A(\lambda)$ as
\begin{align}
D_A (\lambda)&\simeq D_A^{(0)}(\lambda) \left\lbrace 1+ie^{-ik_F\ell}e^{-2ik_F\ell}L_k^{-1-2\beta_{\lambda}}\frac{\Gamma(1+\beta)}{\Gamma(-\beta_{\lambda})} - ie^{ik_F\ell}e^{2ik_F\ell}L_k^{-1+2\beta_{\lambda}}\frac{\Gamma(1-\beta_{\lambda})}{\Gamma(\beta_{\lambda})} \right\rbrace\\
& \equiv D_A^{(0)}(\lambda)(1+\Psi_{\ell}(\lambda)), \nonumber
\end{align} 
where the leading term $D_A^{(0)}(\lambda)$ is the one in Eq. \eqref{DA0} and we introduced the shorthand
\begin{equation}
L_k=4\Big(\ell+\frac{1}{2}\Big)|\sin k_F|.
\end{equation}
The corrections to the scaling are characterised by the difference
\begin{equation}
d_n(\alpha)\equiv \log {Z}_n(\alpha)-\log {Z}_n^{(0)}(\alpha),
\label{dndef}
\end{equation}
that for large $L_k$ can be written as
\begin{equation}
d_n(\alpha)\simeq\frac{1}{2\pi i}\oint d\lambda f_{n,\alpha}(1,x)\frac{d \log[1+\Psi_{\ell}(\lambda)]}{d\lambda}=
\frac{1}{2\pi i}\oint d \lambda f_{n,\alpha}(1,\lambda)\frac{d \Psi_{\ell}(\lambda)}{d\lambda}+\cdots.
\end{equation}
The contour integral can be split as the sum of two contributions  above and below the segment $[-1,1]$
\begin{equation}
d_n(\ell)\simeq\frac{1}{2 \pi i }\left[\int_{-1+i\epsilon}^{1+i \epsilon}-\int_{-1-i\epsilon}^{1-i \epsilon}\right]d \lambda f_{n,\alpha}(1,\lambda)\frac{d\Psi_{\ell}(\lambda)}{d \lambda}.
\end{equation}
The integral is then given by the discontinuity at the branch cut;
the only discontinuous function along this cut is $\beta_{\lambda}$, that we rewrite as (for $-1<x<1$)
\begin{equation}
\beta_{x\pm i\epsilon}=-iw(x)\mp\frac{1}{2}, \quad \quad \mbox{with} \quad \quad w(x)=\frac{1}{2\pi}\log\frac{1+x}{1-x}.
\end{equation}
Hence the discontinuities across the branch cut given by the two terms in $\Psi_{\ell}$ are respectively
\bea
&&\left[ L_k^{-i-2\beta} \frac{\Gamma(1+\beta)}{-\beta}\right]_{\beta=-iw-\frac{1}{2}}-\left[L_k^{-i-2\beta} \frac{\Gamma(1+\beta)}{-\beta}\right]_{\beta=-iw+\frac{1}{2}}\simeq L_k^{2 i w}\gamma(w),
\\
&&\left[ L_k^{-i+2\beta} \frac{\Gamma(1+\beta)}{-\beta}\right]_{\beta=-iw-\frac{1}{2}}-\left[   L_k^{-i+2\beta} \frac{\Gamma(1+\beta)}{-\beta}\right]_{\beta=-iw+\frac{1}{2}}\simeq L_k^{-2 i w}\gamma(-w),
\eea
\begin{figure}[t]
	\includegraphics[width=.31\textwidth]{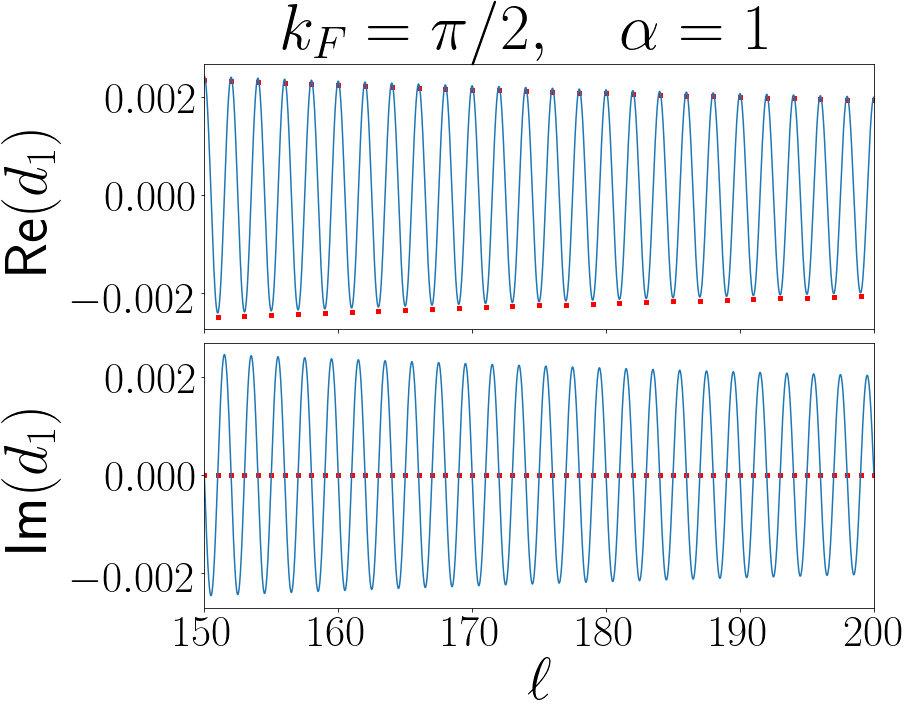}
        \includegraphics[width=.31\textwidth]{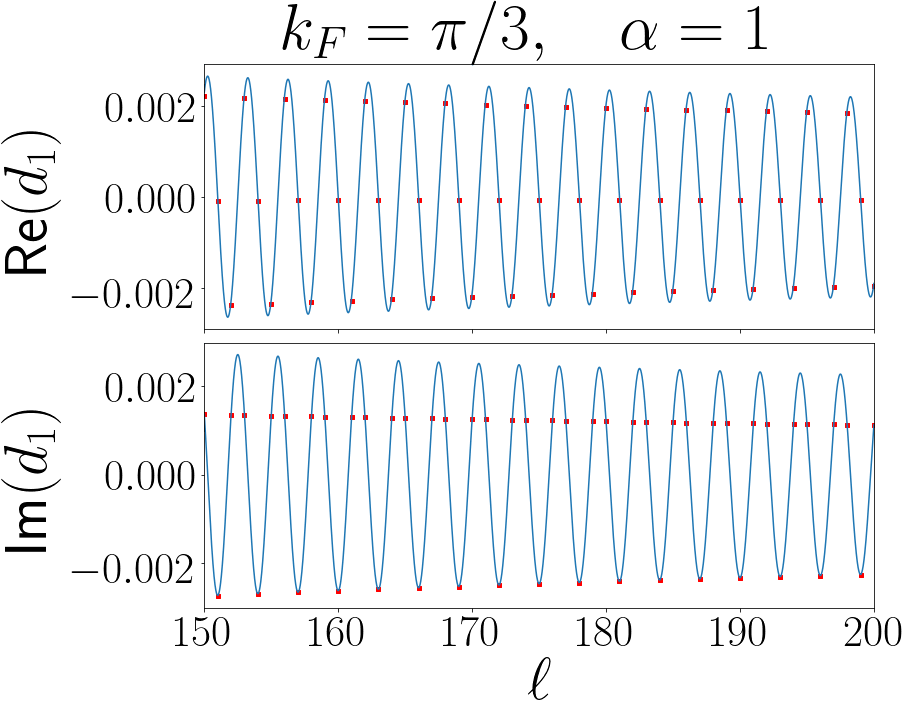}
        \includegraphics[width=.31\textwidth]{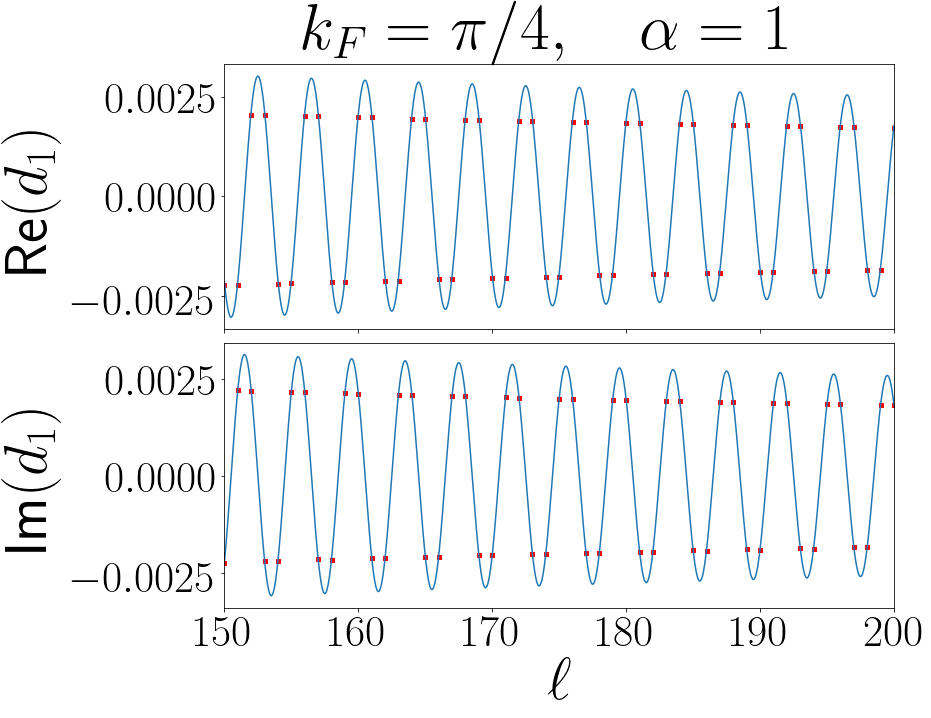}\\
        \includegraphics[width=.31\textwidth]{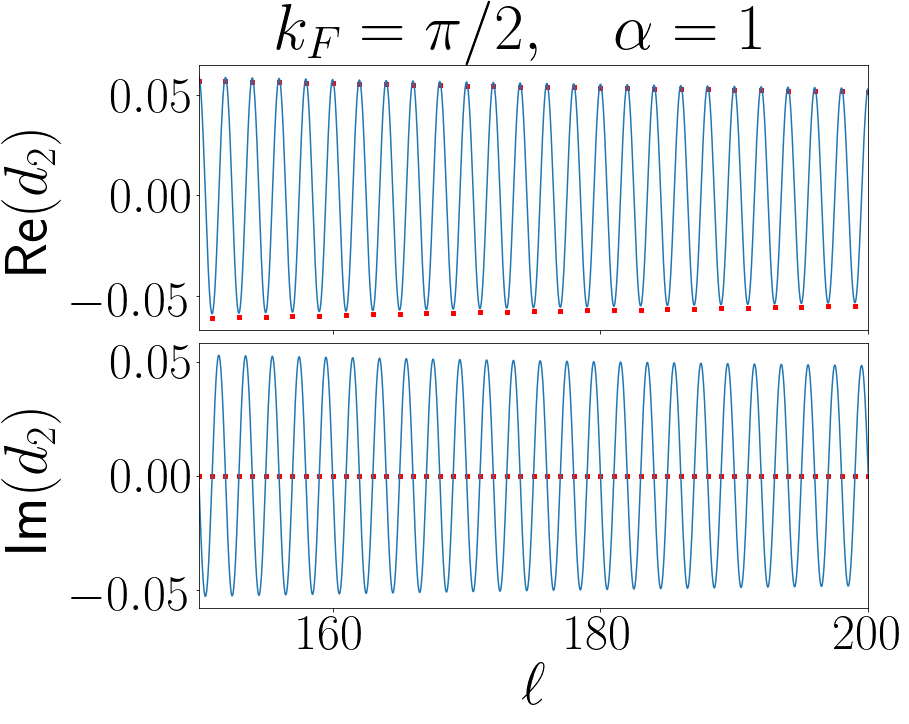}
        \includegraphics[width=.31\textwidth]{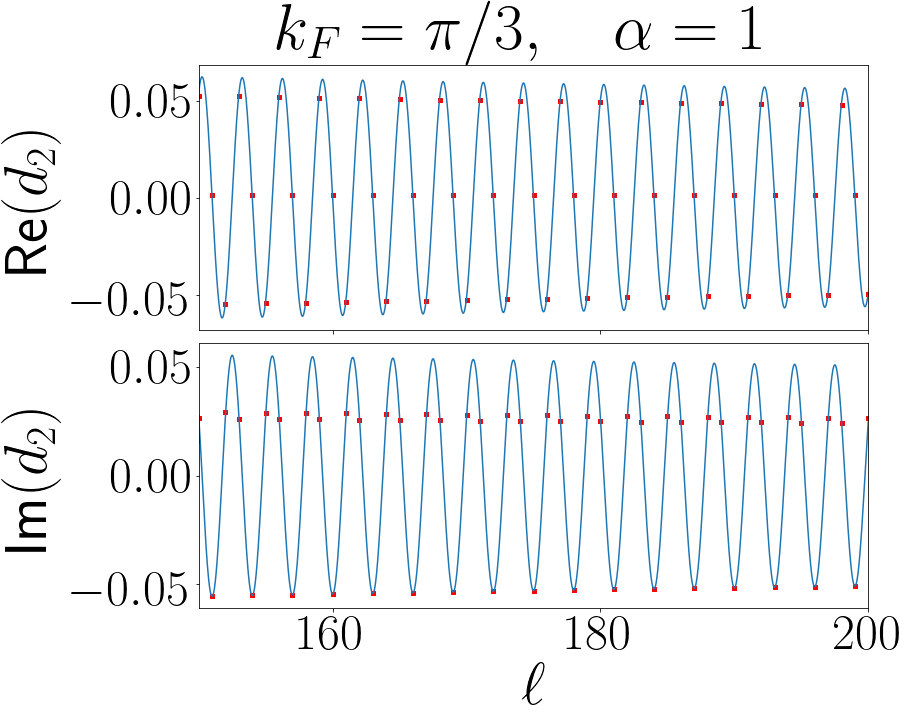}
        \includegraphics[width=.31\textwidth]{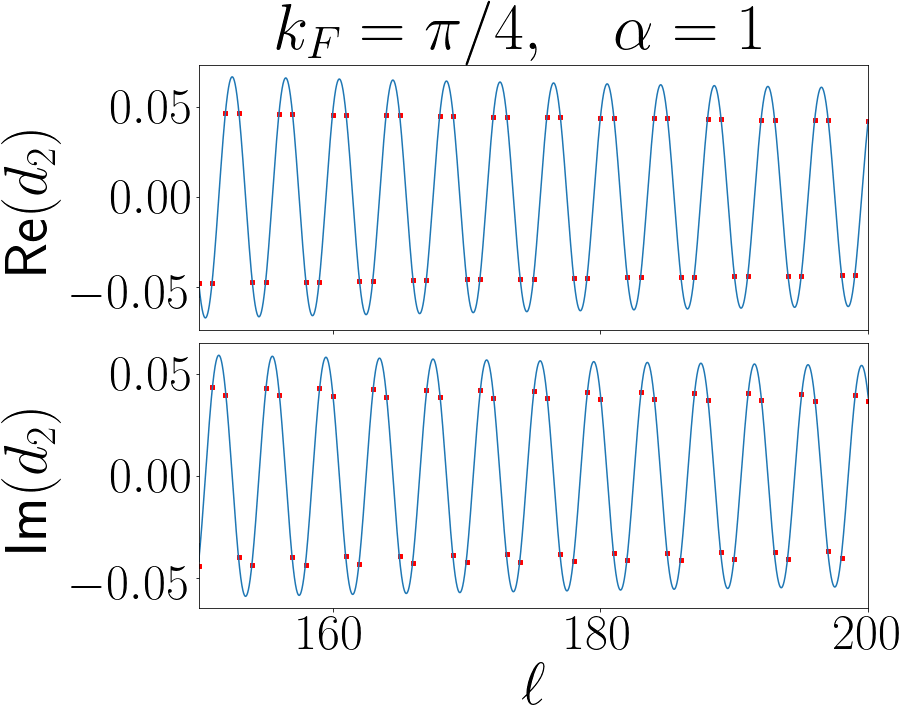}
     \caption{Leading corrections to the scaling for the charged entropies. 
     The plots show the difference $d_n(\alpha)$ defined in Eq. \eqref{dndef}. We focus on $\alpha=1$, $n=1,2$ and different fillings 
     $k_F=\pi/2$  (left), $k_F=\pi/3$ (center) and $k_F=\pi/4$ (right) as function of $\ell$. 
     The numerical data (symbols) match well the calculated leading correction to the scaling \eqref{corrections} from generalised Fisher-Hartwig formula 
     both for real and imaginary parts.}
     \label{Fig11}
\end{figure} 
where we dropped terms of order $O(L_k^{-2})$ compared to the leading ones and defined
\begin{equation}
\gamma(w)=\frac{\Gamma \left( \frac{1}{2} -i w \right)}{\Gamma \left( \frac{1}{2} +i w \right)}.
\end{equation}
We can now perform the change of variable
\begin{equation}
\lambda=\tanh(\pi w), \quad \quad -\infty<w<\infty,
\end{equation}
and integrate by parts using
\begin{equation}
\frac{d}{d w}f_n(\tanh(\pi w),\alpha)=\pi n \left[ \tanh(\pi n w+i \alpha/2) -\tanh(\pi w)\right],
\end{equation}
to finally get
\begin{multline}
d_n(\alpha)\simeq\frac{i n}{2}\int_{-\infty}^{\infty}d w \left( \tanh(\pi w) -\tanh(\pi n w+ i \alpha/2) \right) \times\\
\left[ i e^{-ik_F}e^{-2 i k_F \ell}L_k^{2 i w}\gamma(w)+ i e^{ik_F}e^{2 i k_F \ell}L_k^{-2 i w}\gamma(-w)\right]. 
\end{multline}
This last integral can be performed on the complex plane by residue theorem.
For the first piece of the integral in square brackets, we must close the contour in the upper half plane,  while for the second piece in the lower half plane. 
In principle, we should sum over all residues inside the integration contour, but, since we are interested in the limit of large $L_k$, we can limit ourselves to 
consider the singularities closest to the real axis. 
For the first integral this is at $w= i /(2n) (1-\alpha/\pi)$ while for the second one it is at $w=-i/(2n) (1+\alpha/\pi)$.
Their sum gives
\begin{equation}
\label{corrections}
d_n(\alpha)\simeq i e^{-ik_F (2\ell+1)}L_k^{-\frac{1}{n}\left( 1-\frac{\alpha}{\pi}\right)}\frac{\Gamma\left( \frac{1}{2}+\frac{1}{2n}-\frac{\alpha}{2 \pi n} \right)}{\Gamma\left( \frac{1}{2}-\frac{1}{2n}+\frac{\alpha}{2 \pi n} \right)}- ie^{ik_F (2\ell+1)}L_k^{-\frac{1}{n}\left( 1+\frac{\alpha}{\pi}\right)}\frac{\Gamma\left( \frac{1}{2}+\frac{1}{2n}+\frac{\alpha}{2 \pi n} \right)}{\Gamma\left( \frac{1}{2}-\frac{1}{2n}-\frac{\alpha}{2 \pi n} \right)}.
\end{equation}
This expression represents our final result for the oscillating corrections to the scaling. Let us comment it. 
For $\alpha=0$ the corrections are real and give back the result already found in Ref. \cite{fc-11}. 
Otherwise, Eq. \eqref{corrections} is valid only in the range $-\pi\leq \alpha \leq \pi$ and must be extended periodically outside of it. 
Mathematically, one can found the periodic structure using the entire sum in the generalised Fisher-Hartwig formula \eqref{Drep2o}.
This periodicity in $\alpha$ is identical to what found for PBC in Ref. \cite{fg-20} to which we remand for an extensive discussion. 
Within the principal domain $-\pi\leq \alpha \leq \pi$, the two contributions decay with different power laws (as anticipated above), so that only one of the two is dominating, 
according to the sign of $\alpha$.
However, they become comparable in magnitude as $\alpha$ approaches zero where one should carefully take into account both of them.  
Conversely, for $\alpha$ close to $\pm \pi$, the calculated corrections with $m=\pm1$ become comparable with the leading term ($m=0$) and hence 
a good agreement between Eq. \eqref{corrections} and the data is achieved only at extremely large $\ell$, since in the derivation we assumed that the leading term is
much larger than the corrections.

In Fig. \ref{Fig11} we report the exact numerical data for $d_n(\alpha)$ at fixed $\alpha=1$ and $n=1,2$. 
We observe that, for different fillings, the data always match well the analytic prediction \eqref{corrections} both for real and imaginary part.
Remarkably, the sampling at commensurate points due to the integer values of $\ell$, match well the oscillating structure due to the 
partial fillings.

\subsection{Symmetry resolved moments and entropies}

\begin{figure}[t]
\includegraphics[width=.49\textwidth]{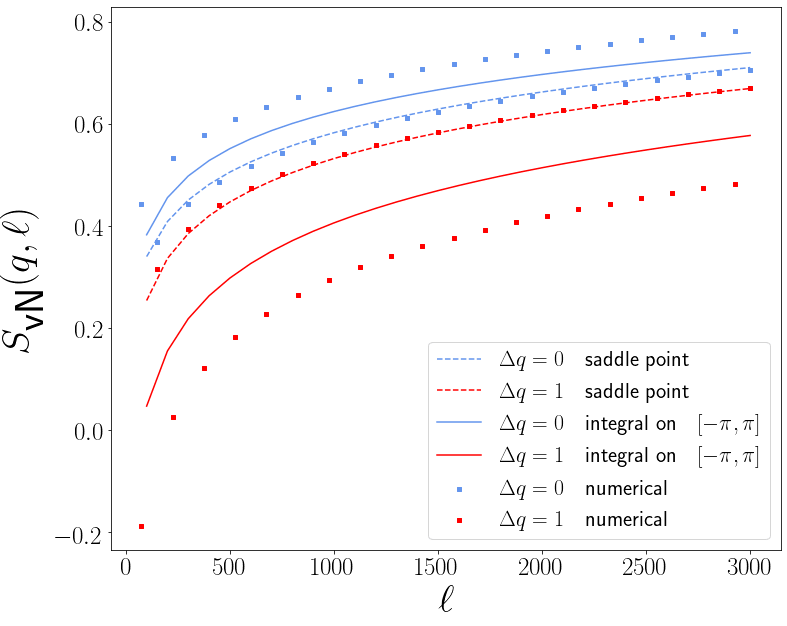}
\includegraphics[width=.49\textwidth]{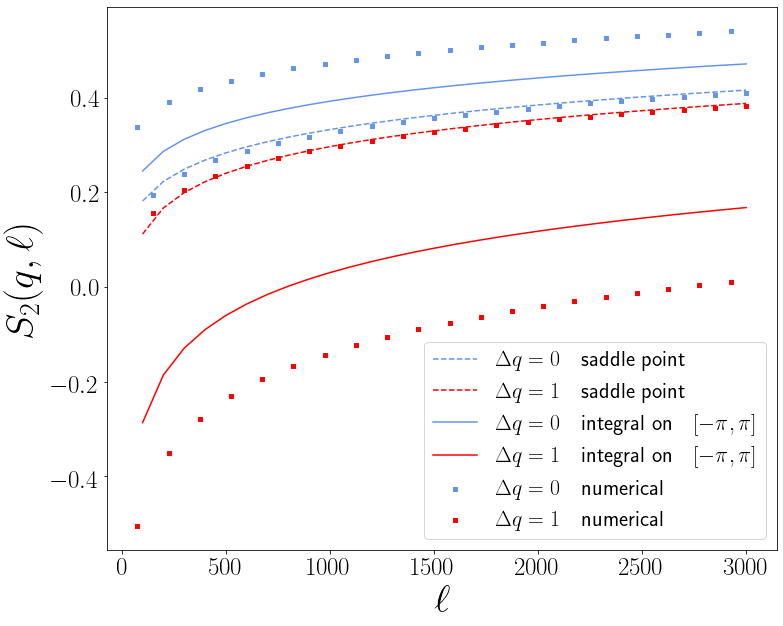}
\caption{Von Neumann (left) and second R\'enyi (right) symmetry resolved entanglement entropies at half filling $k_F=\pi/2$ for one interval of length $\ell$
starting from the boundary. 
The analytical predictions (dashed lines) in Eq. \eqref{eq:SP-SRRE-v2Order} with the amplitudes in Eq. \eqref{bnobc} are compared with the numerical data (symbols) 
for $\Delta q=q-\bar q=0, 1$. 
The agreement of the saddle point results with the numerical data is rather unsatisfactory. 
The observed deviations are caused by the finite domain of the integral in $\alpha$, and are well captured by Eq. \eqref{erf} reported as full lines which falls 
in the middle between the data for even and odd $\ell$. 
}
\label{EntropiesPlots}
\end{figure}

We are finally ready to evaluate the symmetry resolved moments $\mathcal{Z}_n(q)$ as Fourier transform of ${Z}_n(\alpha)$ 
in the saddle point approximation. In this approximation, the leading term is given by $Z_n^{(0)}(\alpha)$ in Eqs. \eqref{ZG} with 
the variance $b_n$ again given by Eq. \eqref{bn}, but with the amplitudes
\begin{equation}
b=\frac12\,, \qquad h_n= -\frac1n\frac1{2\pi^2} \log(4\sin k_F) +\gamma_2(n)\,.
\label{bnobc}
\end{equation}
The symmetry resolved moments and entropies (for large $\ell$) are then given by Eqs. \eqref{eq:SP-FTrans-step3} and \eqref{eq:SP-SRRE-v2Order}, respectively, 
with the amplitudes given in Eq. \eqref{bnobc}.

In Fig. \ref{EntropiesPlots}, we compare the analytical predictions for the symmetry resolved entanglement entropies with the numerical data, focusing on half filling $k_F=\pi/2$.
The agreement is overall rather unsatisfactory, the asymptotic curves look rather close to data for even $\ell$, but too far from those at odd $\ell$.
Although the oscillations are very large, we would have expected the analytic curves to be in the middle of the data, as it happens for the charged moments. 
Actually, the reason of this disagreement is easily identified. Indeed, the asymptotic curve in the saddle point approximation is given by the Fourier transform 
of $Z_n(\alpha)$ where the integral has domain on the entire real axis, rather than the interval $[-\pi,\pi]$. The difference between the two is exponentially small 
in the variance $b_n$, and hence is usually neglected. However, the variance only grows as $\log \ell$; consequently the corrections are algebraic in $\ell$ and so 
rather visible. 
The data at finite, even large, $\ell$ should be better described by the integral on the domain $[-\pi, \pi]$, i.e. 
\beq
{\cal Z}_n(q)
\simeq
Z_n(0)  \int_{-\pi}^\pi \frac{d \alpha}{2\pi} e^{-i \alpha (q-\bar q)- b_n \alpha^2/2}= 
Z_n(0) e^{-\frac{\Delta q^2}{2 b_n}} \frac{{\rm Erf}\big(\frac{i\Delta q+\pi b_n}{\sqrt{2 b_n}}\big)+{\rm Erf}\big(\frac{-i\Delta q+\pi b_n}{\sqrt{2 b_n}}\big)}{2\sqrt{2\pi b_n}},
\label{erf}
\eeq
where ${\rm Erf}(z)$ is the error function. 
The symmetry resolved entanglement entropies obtained, via Eq. \eqref{SvsZ} are reported for $\Delta q=0,1$ in Fig. \ref{EntropiesPlots} as continuous lines 
and, as expected, they fall right in the middle of the numerical data for even and odd $\ell$. 
For extremely large $\ell$, Eq. \eqref{erf} clearly converges to the asymptotic result in Eq. \eqref{eq:SP-FTrans-step3}.
It is instructive to check analytically this slow approach, also to motivate the  strong visible difference compared to the periodic case \cite{bons}.
To this aim, let us focus on the case $\Delta q=0$ which has the smallest corrections. 
The relative difference between the truly asymptotic behaviour and Eq. \eqref{erf} is
\beq
\frac{{\cal Z}_n(0)-  {\cal Z}_n^{\rm asy} (0)}{{\cal Z}_n^{\rm asy} (0)}\simeq {\rm Erfc} \Big(\pi \sqrt{\frac{b_n}2}\Big)\simeq \frac{e^{-b_n \frac{\pi^2}2} }{\sqrt{\pi^3 b_n } }
\sim \frac{\ell^{-\frac{b}{2n}}}{\sqrt{\log \ell}},
\eeq
where ${\rm Erfc}(z)$ is the complementary error function and in the last equality we only used the leading behaviour $b_n\sim b\log\ell$.
It is clear that for OBC these corrections are much more severe than for the infinite system because $b=1/2$ (instead of $b=1$ with PBC). 
For example, for $n=1$, which is the better behaving case among those in Fig. \ref{EntropiesPlots}, we have corrections going like $\ell^{-1/4}$ which are extremely slow.

\section{A general relation between Toeplitz+Hankel matrices and block Toeplitz ones}
\label{sec:proof}

We have seen that in CFT the (charged and neutral) entanglement entropies of one interval in the middle of a semi-infinite chain are related to the ones 
of two intervals in an infinite chain. It is rather natural to wonder whether this correspondence has some akin for free fermions on the lattice. 
This is indeed the case and it is a consequence of a general relation between Toeplitz+Hankel matrices and block Toeplitz ones that 
we are going to show (and that in a similar form was present in Ref. \cite{fc-08}, but for a slightly different class of matrices). 

Let us start from the correlation matrix of two disjoint intervals of equal length $\ell$ at distance $d$ in an infinite system.
We denote this $2\ell\times2\ell$ matrix by $C_2^{\ell,d}$ that is easily obtained from the correlation matrix of a single interval of length $2\ell+d$
given by Eq. \eqref{CApbc} by erasing all rows and column between $\ell+1$ and $\ell+d$. 
It has the block structure
\begin{equation}
C_2^{\ell,d}= \left(\begin{array} {ll}
F& E^{d+\ell}\\ (E^{d+\ell})^T&F
\end{array}\right),
\qquad {\rm with}\quad  F_{i,j}=f_{i-j}, \quad  E_{i,j}^{k}= f_{i-j+k}\,,  
\label{C2}
\end{equation}
and $f_m=\frac{\sin (k_F m)}{\pi m}$ are the elements of the correlation matrix of the single interval \eqref{CApbc}, 
although what follows is true for arbitrary $f_m$ such that $f_m=f_{-m}$.
The four blocks in the matrix $C_2^{\ell,d}$ are Toeplitz matrices, but the total matrix is not. 
(Notice that by rearranging rows and columns it is possible to write $C_2^{\ell,d}$ in such a way that it is a Toeplitz matrix in which the 
elementary block is a $2\times 2$ matrix, a standard form in the literature, see e.g. \cite{fc-08}. 
However this not relevant for us.)

Let us denote by $C_-$ the $\ell\times \ell$ correlation matrix of a block of length $\ell$ at distance $\ell_0$ from the boundary of a semi-infinite system in 
Eq. \eqref{T+H}, which can be written as
 \beq
C_-= F- H^{2\ell_0}\,, \qquad {\rm with}\quad\, H^{2\ell_0}_{i,j}= f_{i+j+2\ell_0}.
\eeq
Let us also introduce the matrix 
 \beq
C_+= F+H^{2\ell_0}\,,
\eeq
which, incidentally, is the correlation matrix of a semi-infinite system with a Neumann boundary condition on the first site $i=1$ (see e.g. \cite{cmv-11b}).
Let us now assume that $\vec v_-$ is a $\ell$-component vector, which is eigenvector of $C_-$, i.e. $C_- \vec v_-=\lambda_{v_-} \vec v_-$. 
Let us introduce the reversed of a vector $\vec v$ as $\vec v^R=(v_\ell, v_{\ell-1},\dots v_1)$. 
Then, the $2\ell$-component vector $\vec V_-\equiv (\vec v_-^R,-\vec v_-)$ is an eigenvector of $C_2^{\ell,2\ell_0+1}$ with eigenvalue $\lambda_{v_-}$, 
as one straightforwardly shows, indeed 
\beq
 \left(\begin{array} {ll}
F& E^{2\ell_0+1+\ell}\\ (E^{2\ell_0+1+\ell})^T&F
\end{array}\right) 
 \left(\begin{array} {c}
\vec v_R\\ -\vec v
\end{array}\right) 
=
 \left(\begin{array} {c}
F \vec v^R- E^{2\ell_0+1+\ell} \vec v\\ (E^{2\ell_0+1+\ell})^T \vec v^R-F \vec v
\end{array}\right), 
\label{p22}
\eeq
and the components of the top vector are
\beq
\sum_{j=1}^\ell (f_{i-j} v_{\ell+1-j} -  f_{i-j+2\ell_0+1+\ell} v_{j})=\sum_{j=1}^\ell  (f_{i-j}  -  f_{i+j'+2\ell_0}) v_{\ell+1-j},
\eeq
which is what we wanted to prove. The same goes on for the bottom vector in Eq. \eqref{p22}.
Similarly, given an eigenvector $\vec v_+$ of $C^+$ of eigenvalue $\lambda_{v_+}$, we have that the vector $\vec V_+\equiv (\vec v_+^R,\vec v_+)$ is 
eigenvector of $C_2^{\ell,2\ell_0+1}$ with eigenvalue $\lambda_{v_+}$.
By construction, all vectors $\vec V_{\pm}$ are orthogonal and hence, since they are $2\ell$, they form a basis.
The main conclusion here is that the spectrum of $C_2$ is the union of the spectra of $C_+$ and $C_-$, a rather remarkable result that likely is known 
in the literature, but we failed to find it (see however \cite{fc-08}). Thus we can generically relate the spectra of block Toeplitz matrices with  Toeplitz+Hankel ones.
We stress that these results are only valid in the fermionic basis and not in the spin one, where the structure is much more complicated because of the 
Jordan-Wigner string \cite{fc-10,ctc15c}.

Let us now move to the main objects of interest here that are the moments $Z_n(\alpha)$, both charged and neutral. 
Since they are just products of functionals of the eigenvalues of the respective correlation matrices, cf. Eq. \eqref{LnChargedMoments}, we have that the moment of 
two intervals of length $\ell$ at distance $2\ell_0+1$ in an infinite chain is the product of the moments built from $C_+$ and $C_-$.
This is valid for arbitrary $\ell$ and $\ell_0$. 
In the scaling limit $\ell,\ell_0\to\infty$, with arbitrary ratio $\ell/\ell_0$, the moments built with $C_+$ and $C_-$ are expected to become equal, 
because both boundary conditions correspond to the same boundary CFT with $g=1$, see e.g.  \cite{fsw-94,cmv-11b}.
Consequently, the scaling form of the moments of one interval at distance $\ell_0$ from the boundary (with both free and Neumann conditions) is the square root of the 
one of two intervals of the same length at distance $2\ell_0+1$ in an infinite system.

\section{Disjoint intervals in the infinite chain}
\label{Sec5}

En route to the calculation of the symmetry resolved entropies for a block away from the boundary, we first discuss the case of two disjoint intervals 
in an infinite system, since we have just seen the two are related.  
In this case, the correlation matrix is given by Eq. \eqref{C2}. 
For the matrices with this structure, a generalisation of the Fisher-Hartwig formula has been conjectured in Ref. \cite{aef-14b} for the evaluation of the total R\'enyi entropies. 
Here, we just state the final result of Ref. \cite{aef-14b} and refer the reader interested into more details to the original reference. 
We write the final result for a  generic piecewise constant symbol $g(\theta)$ with $R$ discontinuities at $\theta=\theta_r$ with $r=1,\dots N$, 
although we are also interested in the case $R=2$ in Eq. \eqref{symbol}. We introduce $t_r\equiv g(\theta)$ with $\theta\in [\theta_{r-1}, \theta_{r})$.
We consider the subsystem $A=[u_1,v_1]\cup [u_2,v_2]$ of total length $|A|=v_1-u_1+v_2-u_2$. 
The final result of Ref. \cite{aef-14b} is 
\begin{equation}
\label{SntwointervalsFH}
S_n(X)=A_n|A|+B_n\log\frac{(v_1-u_1)(v_2-u_2)(v_2-u_1)(u_2-v_1)}{(u_2-u_1)(v_2-v_1)}+ 2 C_n+\dots,
\end{equation}
where the coefficients depend only on the symbol $g(\theta)$ and are given by 
\begin{equation}
A_n= \frac{1}{2 \pi }\int_{-\pi}^{\pi}d\theta f_n(g(\theta)), 
\end{equation}
with $f_n(x)\equiv f_n(x,0)$ in Eq. \eqref{fnxa}, 
\begin{align}
B_n&=2 \sum_{r=1}^R J_n(r,r),\\
C_n&=\sum_{r=1}^{R}I_n(r)- \sum_{1\leq r\neq r'\leq R}\log\left[ 2-2\cos(\theta_r-\theta_{r'}) \right]J_n(r,r'),
\end{align}
where we defined
\begin{align}
J_n(r,r')&=\frac{1}{2 \pi}\int_{t_{r-1}}^{t_r} d \lambda  \frac{df_n(\lambda)}{d \lambda} \omega_{r'}(\lambda), \\
I_n(r)&=\frac{1}{2 \pi i }\int_{t_{r-1}}^{t_r}\frac{d f_n(\lambda)}{d \lambda}\log\left[ \frac{\Gamma\left( \frac{1}{2}-i\omega_r(\lambda)  \right)}{\Gamma\left( \frac{1}{2}+i\omega_r(\lambda)  \right)} \right]d \lambda,
\end{align}
with 
\begin{equation}
 \omega_{r}(\lambda)=\frac{1}{2\pi}\log \Big|\frac{\lambda -t_{r}}{\lambda-t_{r-1}}\Big|.
 \end{equation}
For the specific case of two intervals in the ground state with only two singularities at $\pm k_F$, 
one finally has $A_n=0$, $B_n=1/6(n-1/n) $, and $C_n=\Upsilon_n$. 
This results agrees with the conformal field theory prediction $F_n(x)=1$ \cite{cfh-05}. 

It is completely clear that for the charged entropy of two disjoint intervals, the final result is always given by Eq. \eqref{SntwointervalsFH} with the minor replacement 
$f_n(\lambda)\to f_n(\lambda,\alpha)$. Putting the various pieces together we arrive at
\begin{equation}
\log Z_n(\alpha)=A_{n,\alpha}|A|+B_{n,\alpha}\log\frac{(v_1-u_1)(v_2-u_2)(v_2-u_1)(u_2-v_1)}{(u_2-u_1)(v_2-v_1)}+2 C_{n,\alpha},
\label{Zna2}
\end{equation} 
with 
\begin{align}
A_{n,\alpha}&= \frac{i \alpha k_F}{\pi},\\
B_{n,\alpha}&=\frac{1}{6}\left( n-\frac{1}{n}\right)+\frac{2}{n} \left( \frac{\alpha}{2 \pi}\right)^2 , \\
C_{n,\alpha}&=\Upsilon_n(\alpha).
\end{align}

\section{Block away from the boundary}
\label{6}

In this section, we finally move to the symmetry resolved entropies of a block  $A$ of length $\ell$ placed at distance $\ell_0$ from the boundary (set at $x=0$ for simplicity). 
As proved in Sec. \ref{sec:proof}, the asymptotic scaling behaviour of $Z_n(\alpha)$ is just the square root 
of Eq. \eqref{Zna2} (with $u_2$ and $v_2$ being the mirror images of $u_1=\ell_0$ and $v_1=\ell+2\ell_0$). Hence we  have 
\begin{equation}
\label{LnZOBCdisconnected}
\log Z_n(\alpha)=\frac{i \alpha k_F}{\pi}\ell-\left(\frac{1}{12}\left( n-\frac{1}{n}\right)+\frac{1}{n} \left( \frac{\alpha}{2 \pi}\right)^2  \right) \log\left( \frac{4 \ell^2\ell_0(\ell+\ell_0)}{(2\ell_0+\ell)^2}(2\sin k_F)^2\right) + \Upsilon(n,\alpha)+o(1),
\end{equation}
where $\Upsilon(n,\alpha)$ is the same of Eq. \eqref{Upsilon}. 
The interpretation of the three  terms is the same as in Eq. \eqref{lnZ0res} for the block starting from the boundary. 
This result is valid in the limit $\ell,\ell_0\gg 1$ with their ratio arbitrary and confirms the CFT scaling \eqref{Sadet}.
The only new relevant information is that also for $\alpha\neq 0$, the function $F_{n,\alpha}(x)$ in Eq. \eqref{Sadet} is equal to $1$, as could be likely derived by field theoretical 
means as those in Ref. \cite{cfh-05}. 

\begin{figure}[t]
\includegraphics[width=.49\textwidth]{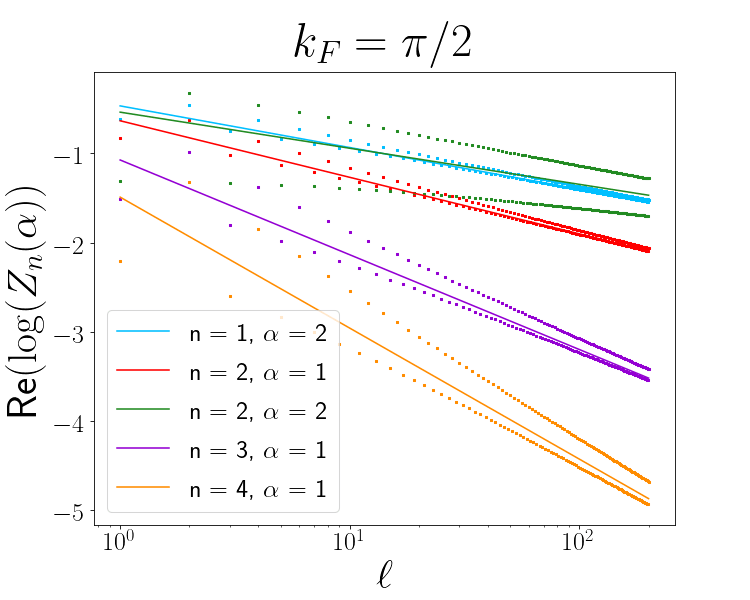}
\includegraphics[width=.49\textwidth]{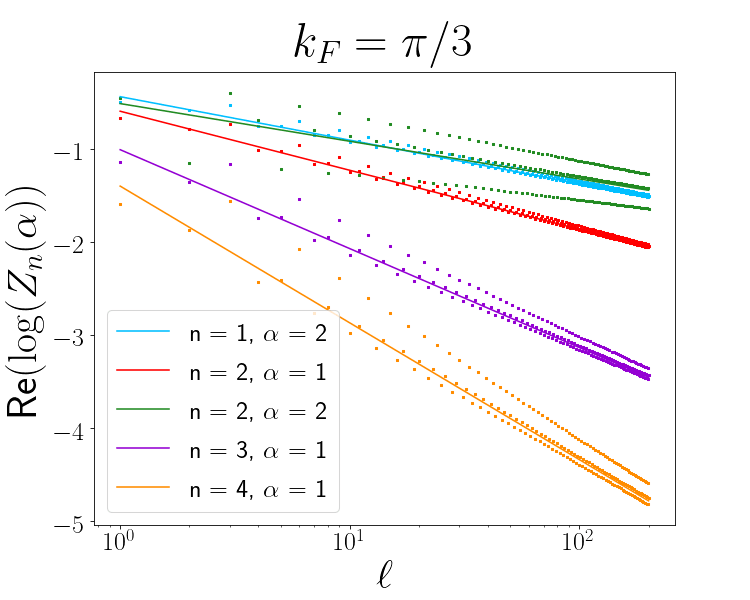}
\caption{Scaling behaviour of the charged R\'enyi entropies $\log Z_n (\alpha)$ for several values of $\alpha$ and $n$.
We fix $\ell_0/\ell=2$ and plot as function of $\ell$.
The numerical data (symbols) are compared with the prediction \eqref{LnZOBCdisconnected}. The agreement is satisfactory, despite the 
presence of strong oscillating corrections.
}
\label{LnZDiscPlot}
\end{figure}

In Fig. \ref{LnZDiscPlot}, we show the comparison between the numerical data and the analytic prediction \eqref{LnZOBCdisconnected} for the R\'enyi entropies in the presence 
of flux for different values of $n$ and $\alpha$, with fillings $k_F=\pi /2$ and $k_F=\pi /3$.
The data are plot as function of $\ell$ at fixed ratio $\ell_0/\ell=2$. 
As expected, the presence of oscillating corrections (with amplitude that increases as  $n$ and $\alpha$ get larger) strongly affect the data, but 
the agreement is satisfactory.

In order to investigate in a better way the correctness of Eq.  \eqref{LnZOBCdisconnected} and to isolate the universal part, it is standard practice to 
construct combinations of entropies of different regions that cancel all non-universal amplitudes (for disjoint intervals the mutual information is one of 
them \cite{fps-09,cct-09}) and 
are explicitly scale invariant, i.e. they depend only on the ratio $\ell/\ell_0$ or, equivalently, on the anharmonic ratio \eqref{anhr}. 
To this aim, let us first define the shorthand $\zeta_A^{(n,\alpha)}$ for $Z_n(\alpha)$ of the subsystem $A$, 
because now we need to refer explicitly to the subsystem. 
The combination of interest is
\begin{equation}
t_n(\alpha)\equiv e^{2i \alpha \langle Q_{[0,\ell_0]} \rangle}
\frac{\zeta_{[\ell_0,\ell_0+\ell]}^{(n,\alpha)}}{\zeta_{[0,\ell_0]}^{(n,\alpha)} \zeta_{[0,\ell_0+\ell]}^{(n,\alpha)} }\,,
\label{tdef}
\end{equation}
where the prefactor is introduced to cancel the mean number of particles in the interval $[0,\ell_0]$. 
The scaling limit of $t_n(\alpha)$  can be written in terms of twist field as 
\begin{equation}
t_n(\alpha)\equiv e^{2i \alpha \langle Q_{[0,\ell_0]} \rangle}
\frac{\langle \mathcal{T}_{n,\alpha} (\ell_0)  \tilde{\mathcal{T}}_{n,\alpha}(\ell_0+\ell)\rangle}{\langle \mathcal{T}_{n,\alpha} (\ell_0) \rangle\langle \mathcal{T}_{n,\alpha} (\ell+\ell_0) \rangle},
\label{tqft}
\end{equation}
in which it is clear that we are taking the ratio of the twist field correlation with its connected part, making it dimensionless and so scale invariant.
Notice in particular that for $\alpha=0$, using also the fact that the entanglement of $A$ is the same as the one of the complement, the above ratio
corresponds to the (exponential) of the R\'enyi mutual information. 

\begin{figure}[t]
\includegraphics[width=.32\textwidth]{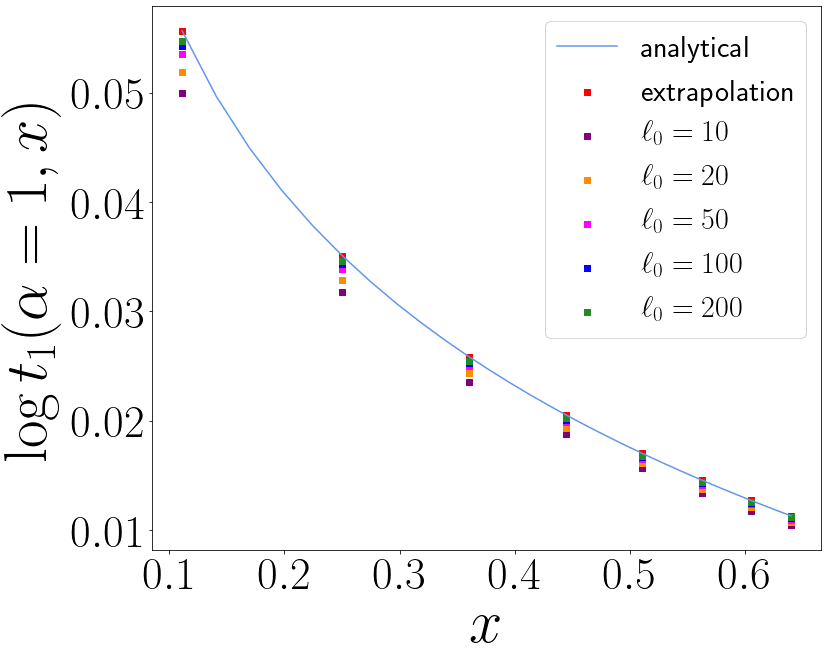}
\includegraphics[width=.32\textwidth]{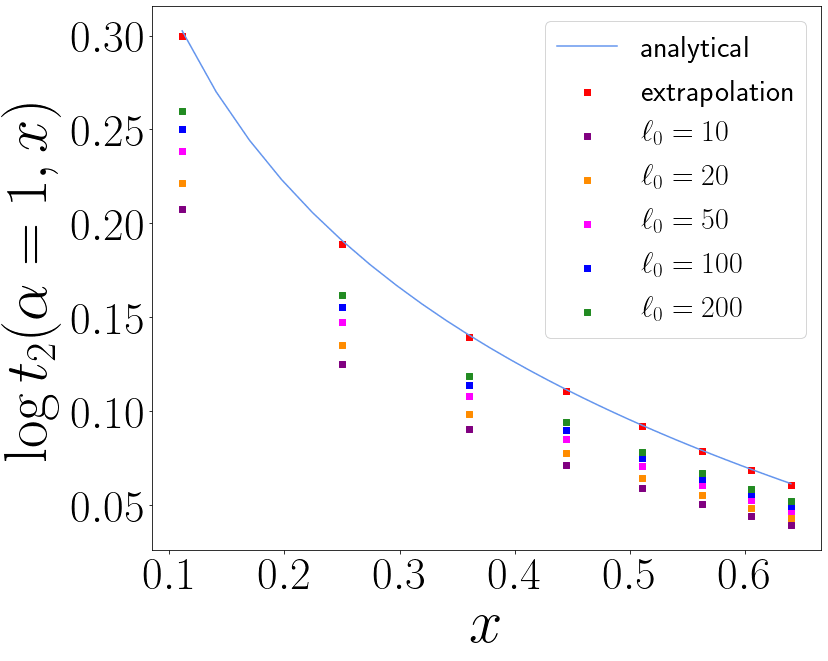}
\includegraphics[width=.32\textwidth]{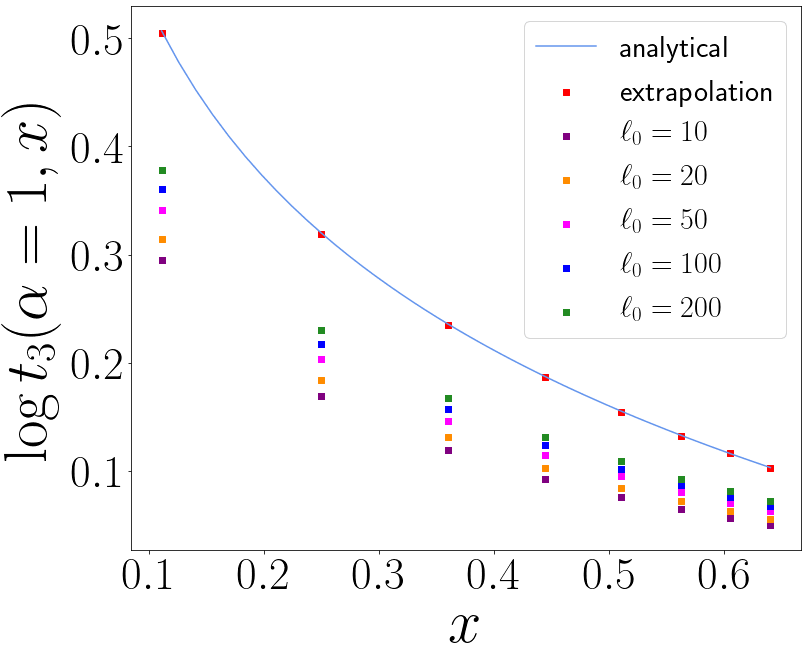}
\caption{
The scale invariant function $\log t_n(\alpha)$ \eqref{tdef} at fixed $\alpha=1$ plotted as function of the anharmonic ratio $x$ for $n=1$ (left), $n=2$ (center) and $n=3$ (right). 
The numerical data are obtained at fixed anharmonic ratio $x$ \eqref{anhr} for increasing values of $\ell_0$ (and $\ell$). 
The data are extrapolated to the scaling limit $\ell,\ell_0\to\infty$ fitting with the first two corrections to the scaling. 
The obtained results match remarkably well our analytic prediction \eqref{RescaledTwist} (continuous line).  
}
\label{RescaledFig}
\end{figure}

Combining Eqs. \eqref{tqft} and \eqref{LnZOBCdisconnected}, it is straightforward to obtain the leading behaviour of $t_n(\alpha)$  
\begin{equation}
\label{RescaledTwist}
\log t_n(\alpha)=
-\left[\frac{1}{12}\left(n-\frac{1}{n} \right)+\frac{1}{n}\left( \frac{\alpha}{2\pi} \right)^2  \right] \log  \frac{\ell^2}{(2\ell_0^2+ \ell)^2} =
-\left[\frac{1}{12}\left(n-\frac{1}{n} \right)+\frac{1}{n}\left( \frac{\alpha}{2\pi} \right)^2  \right] \log  x\,,
\end{equation}
where in the rightmost hand side we recognised the appearance of the anharmonic ratio $x$ as defined in Eq. \eqref{anhr}.
This result is tested against numerics in Fig. \ref{RescaledFig}. It is clear that the date are plagued by scaling corrections.
In order to achieve the scaling limit we work as follows.  For some values of $n$ and $\alpha$ reported in the figure, we computed the ratio
$t_n(x)$ at fixed $x$ and for increasing values of $\ell_0$ from 10 to 200. 
The data present finite $\ell_0$ corrections, which become larger as $n$ and $\alpha$ increase. 
The leading corrections to the scaling behave  $\ell_0^{-(1-\alpha/\pi)/n}$, cf. Eq. \eqref{corrections}.
There are also corrections going as $\ell_0^{-(1+\alpha/\pi)/n}$, $\ell_0^{-(2-\alpha/\pi)/n}$, and $\ell_0^{-1}$ and which one 
is the first subleading depends on $n$ and $\alpha$.
We then perform a fit of the finite $\ell$ data, keeping the first two power-law corrections and extrapolating at $\ell\to\infty$. 
The data obtained following this procedure are reported in Fig. \ref{RescaledFig} and they match extremely well the analytic prediction \eqref{RescaledTwist}.
This is particularly remarkable for $n=2$ and $3$ when the finite $\ell$ data are still very far from their asymptotic values.

\subsection{Symmetry resolution}

For the symmetry resolution of moments and entropies, once again, we exploit the stationary phase approximation which implies the Gaussian approximation \eqref{ZG}
for $Z_n(\alpha)$. 
The parameters defining the variance $b_n$ for large $\ell$, see Eq. \eqref{bn}, in this case read 
\begin{equation}
b=1, \qquad h_n= -  \frac{1}{\pi^2 n}\left[\ln (2 |\sin k_F|)+ \frac12 \log \frac{4\ell_0(\ell+\ell_0)}{(2\ell_0+\ell)^2} \right]+2 \gamma_2(n) . 
\label{bnobc2}
 \end{equation}
 Notice that for $\ell_0\gg\ell$, $h_n$ crosses over to the result in the bulk \eqref{bn2}, as it should. 
The symmetry resolved moments and entropies (for large $\ell$) are then given by Eqs. \eqref{eq:SP-FTrans-step3} and \eqref{eq:SP-SRRE-v2Order}, respectively, 
with the amplitudes given in Eq. \eqref{bnobc2}. 

In Fig. \ref{SymmResDiscFig}, we compare the analytical predictions for the symmetry resolved entanglement entropies with the numerical data, 
focusing on half filling $k_F=\pi/2$ and on $\ell_0/\ell=2$. 
Let us critically compare these results with those with $\ell_0=0$ in Fig. \ref{EntropiesPlots}. 
Especially for $n=2$, we observe that the analytic curves are not in the middle of the data for even and odd $\ell$, but it is slightly shifted. 
The shift is much smaller than the one for $\ell_0=0$ and it can be described by performing the integral on $[-\pi,\pi]$ as in Eq. \eqref{erf}.
Indeed as $\ell\to\infty$, the data crossover  to those for an infinite system, where such an effect was negligible \cite{bons}. 
Indeed for the most relevant case $n=1$, the two curves are extremely close.

\begin{figure}[t]
\includegraphics[width=.5\textwidth]{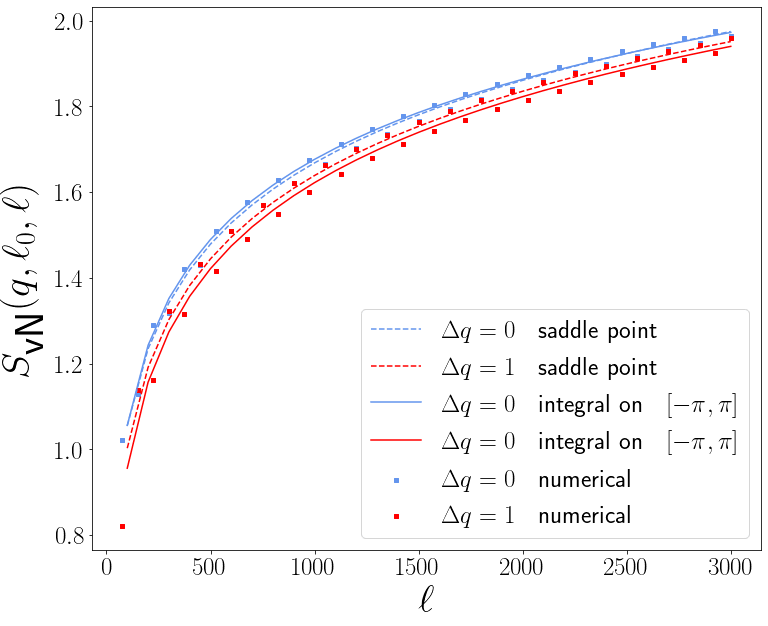}
\includegraphics[width=.5\textwidth]{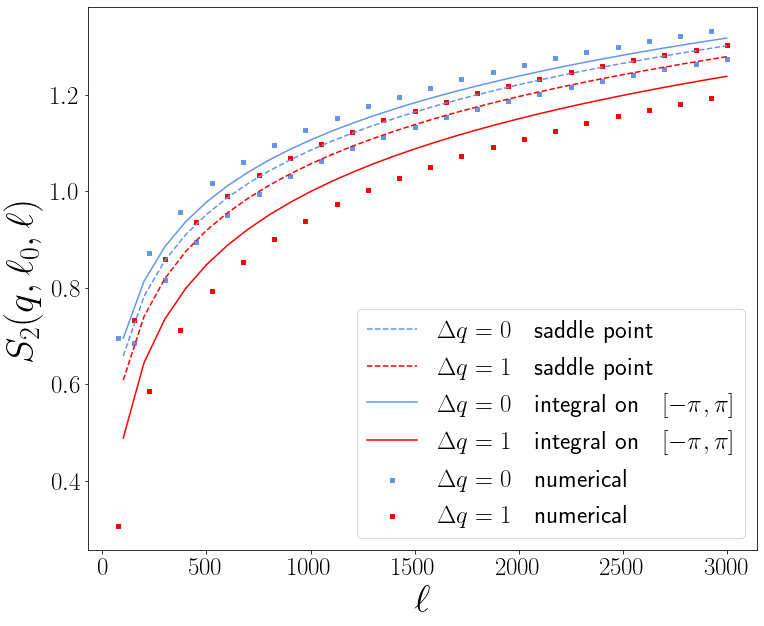}  
     \caption{
Symmetry resolved von Neumann (left) and second R\'enyi (right) entropies at half filling $k_F=\pi/2$ for one interval of length $\ell$
placed at distance $\ell_0$ from the boundary of a semi-infinite chain. We focus on $\ell_0/\ell=2$.
The analytical predictions (dashed lines) in Eq. \eqref{eq:SP-SRRE-v2Order} with the amplitudes in Eq. \eqref{bnobc2} are compared with the numerical data (symbols) 
for $\Delta q=q-\bar q=0, 1$. 
The agreement of the saddle point results with the numerical data is better than the one in Fig. \ref{EntropiesPlots}. However,  
the observed deviations are again caused by the finite domain of the integral in $\alpha$, and are well captured by Eq. \eqref{erf}, full line, which falls 
in the middle between the data for even and odd $\ell$. 
  }
 \label{SymmResDiscFig}
\end{figure}

We conclude this section by studying another universal quantity appearing in the symmetry resolution.
This is related to the variance of the generalised probability distribution function $p_n(q)\equiv \frac{{\cal Z}_n(q)}{{\cal Z}_n}$.
Indeed, as pointed out in Ref. \cite{Luca}, the difference of the variance of a given state with a reference one cancels all non universal factors. 
In our case, we can take the difference of the variance $\langle\Delta q^2  \rangle_n^{\rm OBC}$ for one interval at distance $\ell_0$ from the boundary  
with the one of the infinite system $\langle\Delta q^2  \rangle_n^{\rm PBC}$. 
Since, the variance is just the quadratic term in $Z_n(\alpha)$, it is clear that the 
difference between Eqs. \eqref{bnobc2} and \eqref{bn2} cancels all non-universal and divergent factors, resulting finally in
\begin{equation}
 \label{ExcessVar}
D_n(x) \equiv \langle\Delta q^2  \rangle_n^{\rm OBC}-\langle\Delta q^2  \rangle_n^{\rm PBC}= \frac{1}{2 n \pi^2}\log \frac{4\ell_0(\ell+\ell_0)}{(\ell+\ell_0)^2}= 
\frac{1}{2 n \pi^2}\log (1-x)\,,
 \end{equation}
where in the rightmost hand side we again recognise the anharmonic ratio \eqref{anhr}. 
We can compare the analytic expression of $D_n(x)$ with the numerical results that can be obtained from \cite{Luca}
\begin{equation}
\label{VarNum}
\langle \Delta q^2  \rangle_n=\sum_k \frac{1}{((\nu_k^{-1}-1)^n+1)}-\frac{1}{((\nu_k^{-1}-1)^n+1)^2},
\end{equation}
where the $\nu_k$ are again the eigenvalues of the correlation matrix corresponding to the state and bipartition of interest. 
The excess of variance $D_n(x)$ is then numerically evaluated as the difference between Eq. \eqref{VarNum} evaluated for open and periodic boundary conditions.
The results are shown in Fig. \ref{ExcessVarPlot} for $n=1,2,3$ (respectively from left to right) as function of $x$ at half filling.
The row numerical data at finite $\ell,\ell_0$ are far from the analytic prediction (full line) but tend toward them. 
To be more quantitative, we extrapolate to the scaling limit with a fit that takes into account the first two corrections to the scaling that are expected to go like 
$\ell^{-1/n}$ and  $\ell^{-2/n}$ respectively. These extrapolations are shown in Fig. \ref{ExcessVarPlot} and the agreement is very good, especially when thinking that 
the row data are very far away from the asymptotic ones for $n=2,3$.

\begin{figure}[t]
\includegraphics[width=.32\textwidth]{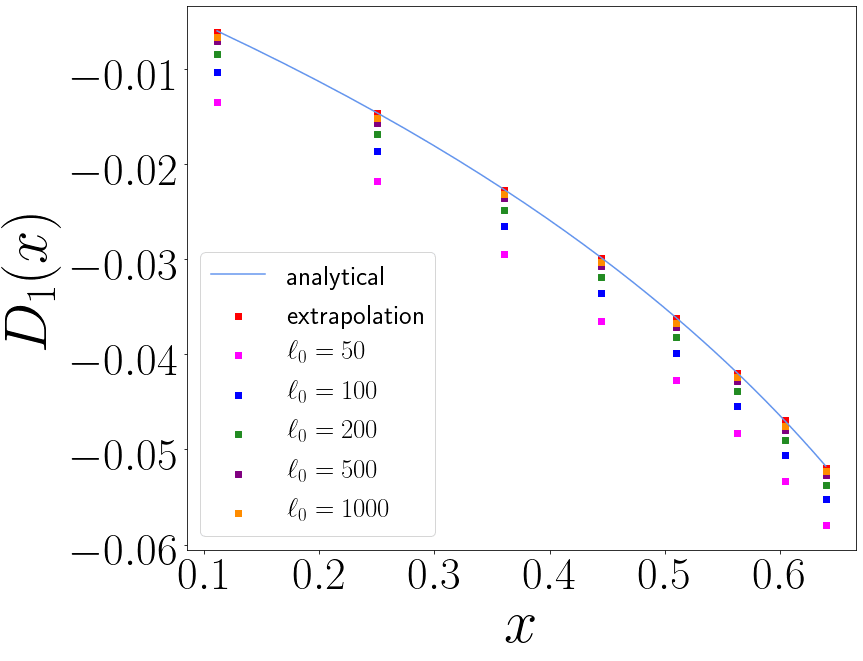}
\includegraphics[width=.32\textwidth]{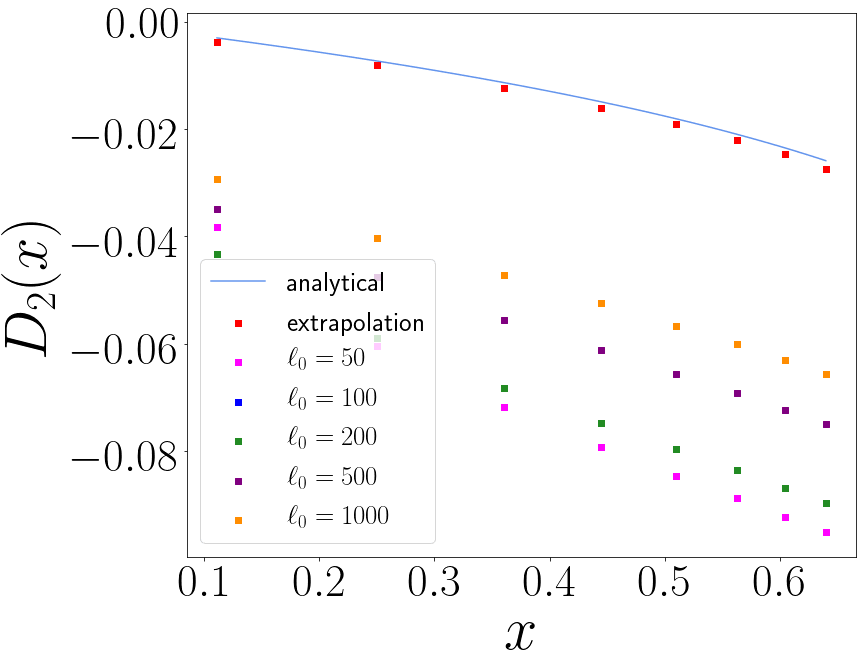}
\includegraphics[width=.32\textwidth]{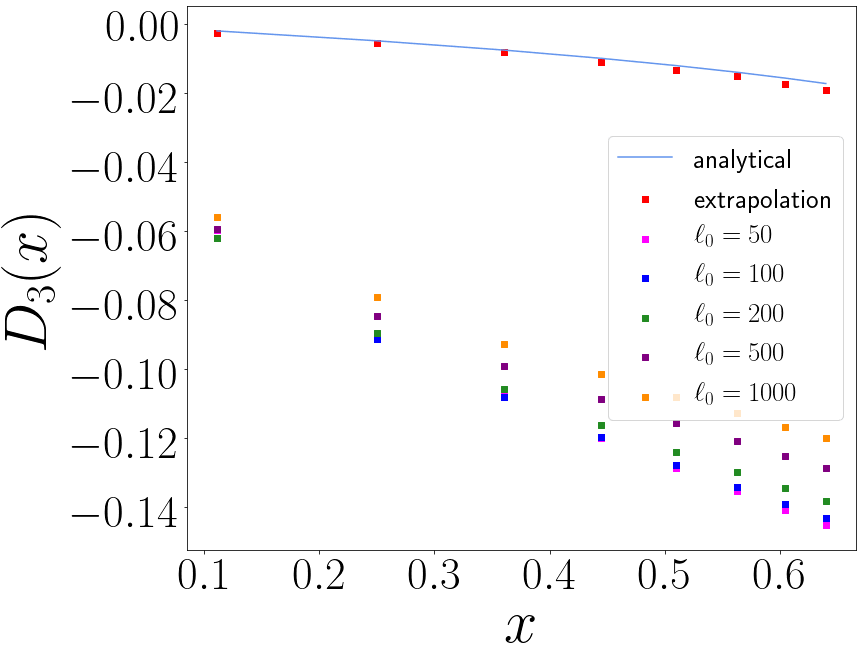}
\caption{Excess of variance  $D_n(x)$ in Eq. \eqref{ExcessVar} as function of $x$ at half filling $k_F=\pi/2$ for $n=1,2,3$ respectively 
obtained from Eq. \eqref{ExcessVar} (blue lines) and from numerical evaluation via correlation matrix (symbols). 
The numerical data are at fixed ratio $x$ for increasing values of $\ell_0$ (and $\ell$). 
The data are extrapolated to the scaling limit $\ell,\ell_0\to\infty$ fitting with the first two corrections to the scaling. 
The obtained results match very well the analytic prediction \eqref{ExcessVar} (continuous line).
}
\label{ExcessVarPlot}
\end{figure}

\section{Conclusions}
\label{7}

In the present work, we characterised the symmetry resolved entanglement entropies in the presence of boundaries. 
We first reported some general results in CFT and then we specialised to spinless free fermions hopping on a 1D lattice. 
We focused on the case of a single block which can be either placed at the boundary or away from it. 
The former case can be treated with rigorous methods based on Fisher-Hartwig formulas \cite{dik-10,fc-11} that allow us to estimate also the corrections to the leading behaviour. 
Conversely, for the interval away from the boundary, we first derived an exact relation with the case of two intervals on the infinite line and then exploited it 
to use a recently proposed conjecture for the latter case \cite{aef-14b}. 
All our analytic results are tested against exact lattice computations. We found that the saddle-point approximation from charged to symmetry resolved entropies 
introduces algebraically decaying corrections in $\ell$ that are much more severe than in the periodic case. 
In all considered instances, we have entanglement equipartition, as expected; however we also identify the first term breaking such an equipartition. 
 
We now briefly mention some possible generalisations and extensions of our results. 
First, thanks to the correspondence between the correlation matrix of the 1D lattice and the overlap matrix of a free Fermi gas \cite{cmv-11,cmv-11b}, our results 
immediately generalise to the latter case, although we did not discuss it here. 
Then, we can exploit our results, to infer predictions for higher dimensional free fermionic lattices and gases with boundaries, using dimensional 
reduction techniques \cite{mrc-20}. Less straightforward generalisations concern instead the effect of the boundaries in the symmetry resolved massive field 
theory. For integrable field theories, one should join the boundary approach of Ref. \cite{cd-09b} with the form factors of the composite twist fields \cite{hc-20}.
Instead for free massive theories, we should generalise the techniques of Refs. \cite{cfh-05,mdc-20b} to the presence of boundaries.

\section*{Acknowledgements}
We are extremely grateful to Sara Murciano for very useful discussions and for comparison of some of the reported results. 
We also thank Luca Capizzi and Paola Ruggiero for useful discussions and collaboration on related topics. 
Both authors acknowledge support from ERC under Consolidator grant number 771536 (NEMO).

\end{document}